%% file: main.tex
\newif\ifanonymous
\newif\ifdraft

\anonymousfalse
\draftfalse
\InputIfFileExists{localflags}{}

\documentclass[ conference]{IEEEtran}

\ifdraft
\fi

\usepackage{versions}

\includeversion{full}
\excludeversion{marked}
\excludeversion{conf}
\usepackage[printonlyused]{acronym}
\usepackage{amsmath}
\usepackage{amsthm}

\usepackage{color}
\usepackage{caption}

\usepackage[operators, adversary, advantage, probability, lambda, mm, sets, asymptotics]{cryptocode}
\let\labelindent\relax
\usepackage[inline]{enumitem} 
\usepackage{float}
\usepackage{fmtcount} 
\usepackage{hyperref}
\usepackage[capitalise]{cleveref}
\usepackage{graphicx} 
\usepackage[utf8]{inputenc}
\usepackage{msc}
\usepackage{listings} 
\usepackage{nccmath} 
\usepackage{stmaryrd}

\usepackage{xspace}
\usepackage{tabularx}
\usepackage{todonotes}
\usepackage{booktabs} 
\usepackage[framemethod=tikz]{mdframed}
\usepackage{etoolbox}
\usepackage[final]{pdfpages}

\AfterEndEnvironment{lstlisting}{\par}

\usepackage{titlesec}
\makeatletter
\def\paragraph{\@startsection{paragraph}{4}%
  \z@\z@{-\fontdimen2\font}%
  {\normalfont}}
\makeatother

\usepackage{bookmark}
\bookmarksetup{
  open,
  numbered,
  depth=5, 
}
\setdescription{itemsep=4pt} 

\graphicspath{ {./img/} }
\usepackage{multirow}

\usepackage[style=ieee,backend=biber,sorting=none,firstinits=true]{biblatex}
\input{bibcleanup}

\input{macros-basic}

\input{macros-meta}

\input{macros-fulltricks}

\input{macros-tikz}
\input{macros-listings}

\input{macros-listings-sapic}
\input{macros-listings-json}

\input{macros-enumitem}

\input{macros-local}

\input{macros-revision.tex}

\begin{document}

\ifanonymous
\author{}

\else
\author{
    \IEEEauthorblockN{Ilkan Esiyok\IEEEauthorrefmark{1},
        Pascal Berrang\IEEEauthorrefmark{2},
        Katriel Cohn-Gordon\IEEEauthorrefmark{3},
        Robert K\"{u}nnemann\IEEEauthorrefmark{1}  }
    \IEEEauthorblockA{\IEEEauthorrefmark{1}CISPA Helmholtz Center for Information Security \{ilkan.esiyok, robert.kuennemann\}@cispa.de}
    \IEEEauthorblockA{\IEEEauthorrefmark{2}University of Birmingham and Nimiq \{p.p.berrang@bham.ac.uk\}}
    \IEEEauthorblockA{\IEEEauthorrefmark{3}Meta \{me@katriel.co.uk\}}
    }

\fi


\title{Accountable JavaScript Code Delivery}

\begin{marked}
    \onecolumn
    \input{summaryofchanges.tex}
        \newpage
        \vfill
	\begin{center}
            {\huge Revised version with
            differences marked in \fixed{red} }
	\end{center}
        \vfill
        \newpage
\twocolumn	
\end{marked}


\label{main:paper}
\maketitle
    \begin{abstract}
        \input{abstract}
    \end{abstract}


\input{intro}
\input{background}
\input{use-cases}

\input{approach}

\input{manifest-intro}

    \input{manifest-order}
    \input{manifest-trust}
    \input{manifest-types}
    \input{manifest-sandboxing}
\input{case-studies}

\input{measurement}
\input{sxg}

\input{protocol}

    \input{verification-intro}

    \input{code-verify}
    \input{transparency-log-intro}
    \input{transparency-logs}
\input{eval}

\input{limitations}
\input{related}

\input{discussion}
\input{conclusion}

\printbibliography

\begin{appendices}
\renewcommand{\thesectiondis}[2]{\Alph{section}:}
    \input{verification}

    \input{claim-verification}
    \begin{full}
        \input{eval-detail}

    \end{full}
    \input{acronyms}
\end{appendices}

\begin{marked}
        \onecolumn
        \section{Revised version (unmarked)}
        \vfill
	\begin{center}
            {\huge Revised version (unmarked) }
	\end{center}
        \vfill
        \newpage
        \includepdf[pages=-]{accjs-conf.pdf} 
        \section{Old version }
        \vfill
	\begin{center}
            {\huge Old Version }
	\end{center}
        \vfill
        \newpage
        \includepdf[pages=-]{pdf/accjs-conf-ndss23-r0.pdf}
\end{marked}

\end{document}

%% file: bibcleanup.tex
\AtEveryBibitem{
 \clearlist{address}
 \clearfield{eprint}
 \clearfield{isbn}
 \clearfield{issn}
 \clearlist{location}
 \clearfield{month}
 \clearfield{date}
 \clearfield{doi}
 \ifentrytype{book}{
  \clearname{editor}
  }{}
\ifentrytype{inproceedings}{

  \clearname{editor}
  \clearfield{volume}
  \clearfield{url}
  \clearfield{eventtitle}
  \clearlist{publisher}
  \clearfield{pages}
  \clearfield{numpages}
  \clearfield{series}

}{}
\ifentrytype{incollection}{
  \clearname{editor}
  \clearfield{volume}
  \clearfield{url}
  \clearlist{publisher}
  \clearfield{series}
  \clearfield{organization}
  \clearfield{institution}
  \clearfield{pages}
}{}
\ifentrytype{proceedings}{
 \clearname{editor}
 \clearfield{volume}
 \clearfield{url}
}{}
\ifentrytype{article}{
 \clearfield{url}
 \clearfield{pages}
 \clearlist{publisher}
 \clearfield{series}
 \clearfield{volume}
}{}
 \ifentrytype{misc}{ 
  \clearname{editor}
  \clearfield{howpublished}
  \clearlist{publisher}
  \clearfield{organization}
  \clearfield{institution}
  \clearlist{year}
  \clearfield{urlyear} 
  }{}
}

%% file: macros-basic.tex
\makeatletter
\DeclareRobustCommand{\defeq}{\mathrel{\rlap{%
  \raisebox{0.3ex}{$\m@th\cdot$}}%
  \raisebox{-0.3ex}{$\m@th\cdot$}}%
  =}
\DeclareRobustCommand{\eqdef}{=\mathrel{\rlap{%
  \raisebox{0.3ex}{$\m@th\cdot$}}%
  \raisebox{-0.3ex}{$\m@th\cdot$}}%
  }
\makeatother

\newtheorem{theorem}{Theorem}

\newtheoremstyle{mycase}{}{}{}{}{\bf}{.}{.5em}{\thmnote{#1:}~\normalfont
  #3}
\theoremstyle{mycase}

\numberwithin{subcase}{mycase}

\newtheoremstyle{component}{}{}{}{}{\itshape}{.}{.5em}{\thmnote{#3}#1}
\theoremstyle{component}

\newcommand{\ledot}{\mathrel{\ooalign{\hss\raise.200ex\hbox{$\cdot$}\hss\cr$\le$}}}
\newcommand{\gedot}{\mathrel{\ooalign{\hss\raise.200ex\hbox{$\cdot$}\hss\cr$\ge$}}}

\newcommand{\rdraw}{\ensuremath{\stackrel{\,\,\mathcal{R}}{\leftarrow}\xspace}}

\newcommand{\pk}{\ensuremath{\mathit{pk}}\xspace}
\newcommand{\sk}{\ensuremath{\mathit{sk}}\xspace}

%% file: macros-meta.tex
\newcommand{\mathcmd}[1]{{\normalfont\ensuremath{\operatorname{#1}}}\xspace}

\newcommand{\mathfun}[1]{\mathcmd{\mathit{#1}}}

\newcommand{\mathvalue}[1]{\mathcmd{\mathit{#1}}}

\newcommand{\textop}[1]{\relax\ifmmode\mathop{\text{#1}}\else\text{#1}\fi}

%% file: macros-fulltricks.tex
\newcommand{\fullversionref}{full-version}

\newcommand{\theappendixorfull}[1]{%
\processifversion{conf}{the full version~\cite[Appendix~\ref{F-#1}]{\fullversionref}}%
\processifversion{full}{Appendix~\ref{#1}}%
        }

\newcommand{\fcite}[1]{\processifversion{full}{~\cite{#1}}}

%% file: macros-tikz.tex
\usetikzlibrary{arrows,shapes,positioning,shadows,trees,automata,tikzmark,fit,backgrounds}

\tikzset{
	box/.style       = {draw, drop shadow, rectangle, rounded corners=2pt, thick, fill=white},
	basic/.style       = {font=\Large, align=center},
	verifier/.style    = {basic},
	user/.style          = {basic},
	issuer/.style         = {basic},
	website/.style         = {basic},
	ledger/.style          = {basic},
	group/.style       = {draw, fill=lightgray!50, very thick,draw=gray, ellipse, sloped,inner sep=2pt},
	label/.style       = {font=\small\sffamily, text centered},
	edgel/.style
        = {midway,sloped,above,font=\small,rectangle,fill=white},
}

%% file: macros-listings.tex
\crefname{lstlisting}{Listing}{Listings}
\Crefname{lstlisting}{Listing}{Listings}

%% file: macros-listings-sapic.tex
\lstdefinelanguage{sapic}{
  morekeywords=[1]{
	out, in, if, then, else, event, insert, delete, lookup, as, in, lock, unlock, let
    },
  morecomment=[l]{\#},
  morecomment=[l]{//},
  morestring=[b]',
  morestring=[s]{`}{'},
}

\lstdefinestyle{sapic}{
    language={sapic}, 
	stringstyle=\rm\textup,
	keywordstyle=[1]\sf,
 	keywordstyle=[2]\rm\upshape,
	literate={||}{{$\mid$}}1
	{new}{{$\nu$}}1
	{<}{{$\langle$}}1
	{>}{{$\rangle$}}1
	{(}{{$($}}1
	{)}{{$)$}}1
}

%% file: macros-listings-json.tex
\colorlet{punct}{red!60!black}
\definecolor{delim}{RGB}{20,105,176}
\colorlet{numb}{magenta!60!black}
\lstdefinelanguage{json}{
    literate=
     *{:}{{{\color{punct}{:}}}}{1}
      {,}{{{\color{punct}{,}}}}{1}
      {\{}{{{\color{delim}{\{}}}}{1}
      {\}}{{{\color{delim}{\}}}}}{1}
      {[}{{{\color{delim}{[}}}}{1}
      {]}{{{\color{delim}{]}}}}{1},
}

%% file: macros-enumitem.tex
\newlist{stepenum}{enumerate}{1} 
\setlist[stepenum]{label=\arabic*., ref=\numberstringnum{\arabic*}, leftmargin=*}
\crefname{stepenumi}{step}{steps}

%% file: macros-local.tex
\newcommand{\nonce}{\mathvalue{n}}
\newcommand{\version}{\mathvalue{v}}
\newcommand{\textURL}{\text{URL}}
\newcommand{\ltextURL}{\text{url}}

\newcommand{\com}{\mathvalue{com}}

\newcommand{\HTML}{\text{HTML}}
\newcommand{\sigtuple}{\mathvalue{sig}_W}
\newcommand{\siglog}{\mathvalue{sig}_L}

\newcommand{\Http}{\text{HTTP}}

\newcommand{\measurementInSignature}{\mathit{\varphi}}

\newcommand{\CodeStapling}{\mathfun{CodeStapling}}
\newcommand{\CodeDelivery}{\mathfun{CodeDelivery}}
\newcommand{\Sign}{\mathfun{sign}}
\newcommand{\VerifySign}{\mathfun{verif}}
\newcommand{\Measure}{\mathfun{measure}}

\newcommand{\accjsheader}{\texttt{x-acc-js-link}\xspace}
\newcommand{\accjsman}{\texttt{x-acc-js-man}\xspace}

\newcommand{\atup}[1]{\langle #1 \rangle}

\newcommand{\Corrupted}{\mathit{Corrupted}\xspace}
\newcommand{\PAccept}{\mathit{PAccept}\xspace}
\newcommand{\LearnedFromW}{\mathit{WSend}\xspace}

\newcommand{\DUploads}{\mathit{DUploads}\xspace}
\newcommand{\CExecutes}{\mathit{CExec}\xspace}
\newcommand{\Logged}{\mathit{Log}\xspace}
\newcommand{\CRecent}{\mathit{CRecent}\xspace}
\newcommand{\KU}{\mathit{KU}\xspace}

\newcommand{\manseq}{\mathvalue{seq}}

\newcommand{\mantype}{\mathvalue{type}}
\newcommand{\mantrust}{\mathvalue{trust}}

\newcommand{\mansrctype}{\mathvalue{src\_type}}
\newcommand{\mansrc}{\mathvalue{src}}
\newcommand{\mansrcdoc}{\mathvalue{srcdoc}}
\newcommand{\mansandbox}{\mathvalue{sandbox}}

\newcommand{\mancrossorigin}{\mathvalue{crossorigin}}

\newcommand{\sid}{\mathvalue{sid}}
\newcommand{\ts}{\mathvalue{ts}}

\newcommand{\inline}{\mathvalue{inline}}
\newcommand{\eventhandler}{\mathvalue{event\_handler}}

\newcommand{\external}{\mathvalue{external}}
\newcommand{\iframe}{\mathvalue{iframe}}
\newcommand{\mathjavascript}{\mathvalue{javascript}}
\newcommand{\assert}{\mathvalue{assert}}
\newcommand{\blindtrust}{\mathvalue{blind-trust}}
\newcommand{\delegate}{\mathvalue{delegate}}

\newcommand{\script}{\mathvalue{script}}

\newcommand{\dynamic}{\mathvalue{dynamic}}
\newcommand{\sync}{\mathvalue{sync}}
\newcommand{\async}{\mathvalue{async}}
\newcommand{\defer}{\mathvalue{defer}}

\newcommand{\persistent}{\mathvalue{persistent}}

\newcommand{\documentwrite}{\mathvalue{document\_write}}

\newcommand{\gensxg}{\mathvalue{gen-signedexchange}}

\usepackage{xargs}
\newcommandx{\todoP}[2][1=]{\todo[inline,linecolor=blue,backgroundcolor=blue!25,bordercolor=blue,#1]{P: #2}}
\newcommandx{\todoI}[2][1=]{\todo[inline,linecolor=green,backgroundcolor=green!25,bordercolor=green,#1]{#2}}
\newcommandx{\todoR}[2][1=]{\todo[inline,linecolor=yellow,backgroundcolor=yellow!25,bordercolor=yellow,#1]{R: #2}}
\newcommandx{\todoK}[2][1=]{\todo[inline,linecolor=red,backgroundcolor=red!25,bordercolor=red,#1]{K: #2}}

\newcommand{\cmark}{\ding{51}}%
\newcommand{\xmark}{\ding{55}}

\newcommand{\generateManifest}{\ensuremath{\mathtt{generate\_manifest}}}

\definecolor{cites}{RGB}{130,147,86}  
\definecolor{links}{RGB}{16,120,150}    
\makeatletter
\hypersetup{
  colorlinks   = true,  
  urlcolor     = blue,  
  linkcolor    = links, 
  citecolor    = cites, 
  final        = true,
}
\makeatother


%% file: macros-revision.tex
\usepackage{xcolor}
\definecolor{darkred}{rgb}{0.8,0,0}
\definecolor{ferrarired}{rgb}{1.0, 0.11, 0.0}

\newcounter{reviewer}
\newcounter{comment}[reviewer]

\ifdraft%
\newenvironment{nnew}{\color{darkred}}{}
\newcommand{\fixed}[1]{{\color{darkred}#1}}
\def\reviewnote#1{\marginpar{\footnotesize#1}}
\else%
\newenvironment{nnew}{}{}
\newcommand{\fixed}[1]{#1}
\def\reviewnote#1{}
\fi

%% file: abstract.tex
The Internet is a major distribution platform for web applications, but there are
no effective transparency and audit mechanisms in place for the web. Due to the ephemeral nature
of web applications, a client visiting a website has no guarantee that the
code it receives today is the same as yesterday, or the same as other visitors
receive. Despite advances in web security, it is thus challenging to audit
web applications before they are rendered in the browser. We propose
Accountable JS, a browser extension and opt-in protocol for accountable
delivery of active content on a web page. We prototype our protocol,
formally model its security properties with the \textsc{Tamarin} Prover, and
evaluate its compatibility and performance impact with case studies
including WhatsApp Web, AdSense and Nimiq. 

Accountability is beginning to be deployed at scale, with Meta’s recent announcement of Code Verify 
available to all 2 billion WhatsApp users, but there has been little formal analysis of such protocols.
We formally model Code Verify using the \textsc{Tamarin} Prover and compare its properties to our Accountable JS protocol. We also compare Code Verify’s and Accountable JS extension's performance impacts on WhatsApp Web.

%% file: intro.tex
\section{Introduction}\label{sec:intro}

Over the years, the web has transformed from an information system
into a decentralised software distribution platform.  Websites
are programs that are freshly fetched whenever accessed and the web
browsers are runtime environments.
This design implies that when a user opens a website, they have no reason to
trust it will run the same program that it did yesterday or the same
program that other users receive. 
Instead, the application loaded may vary over time, and different users may receive
different codes.

The majority of web pages, and even web applications, have neither specified
security goals nor the need to establish them. Nevertheless, for some websites,
maintaining trust between developers and users is part of the
business model:
\begin{itemize}[leftmargin=1.2em]
    \item a private email provider might wish to reassure users that it will always encrypt their messages,
    \item a cryptocurrency wallet might wish to guarantee that it has no access to users' funds, or
    \item a tracking pixel might wish to prove that it only receives data that is explicitly sent to it.
\end{itemize}
Some academic proposals for secure protocols implemented for browsers include
TrollThrottle~\cite{esiyokTrollThrottleRaisingCost2020} and JavaScript
Zero~\cite{schwarzJavaScriptZeroReal2018};
\begin{nnew}
    industry proposals include
payment platforms such as
\href{https://stripe.com}{Stripe} and
\href{https://squareup.com/}{Square},
chat protocols such as
\href{https://web.whatsapp.com/}{WhatsApp Web},
\href{https://www.messenger.com/}{Facebook Messenger}
\fixed{\reviewnote{R\ref{r4}.\ref{r4:examples}}}
and
\href{https://hydrogen.element.io/}{Matrix's Hydrogen client},
encrypted cloud storage such as 
\href{https://mega.io/}{MEGA}
or 
\href{https://spideroak.com/}{SpiderOak}.
\end{nnew}
A concrete example is \href
{https://www.nimiq.com}{Nimiq}, an entirely web-based digital
currency managing private keys in the browser.
It is challenging for such websites to make verifiable guarantees to their users:
a compromised or malicious web server
can precisely target classes of users: the email provider might
disable encryption on a specific IP range, the cryptocurrency wallet
might redirect payments made in some countries, or the tracking pixel
might exfiltrate data only for certain users.

\paragraph{Auditing} 
A common risk mitigation strategy is \emph{auditing}: a developer who wishes
to build trust appoints external auditors to inspect the client code. This can
include both vulnerability research (e.g. via bug bounties) or commissioned
security audits. Audits work well where it is possible for a user to verify
that the code they are running is the same code that was audited, for example
when binaries are received via third party package repositories or app stores
that control the distribution and targeting. App stores do not usually permit
developers to deliver different codes to different users for the same app,
except in a restricted set of circumstances such as for beta testing new
features.

However, auditing does not work for web applications: a compromised or
malicious web server can simply choose at load time to deliver 
unaudited code to a user. No matter how careful the audit or even verification
of the web application, users cannot know that they are receiving the audited
code. Large
parts of modern web security thus depend on techniques like sandboxing or
access control to critical resources like cameras, but fail to capture
properties defined in the context of the application (e.g. authorisation of
transactions in a payment system).

\paragraph{Accountability} 
A second risk mitigation strategy is \emph{accountability}, 
where developers can be held accountable for applications
which they publish. In curated software repositories such as Debian
GNU/Linux or the Apple App Store, developers' code is reviewed and malicious
or compromised code is linked to their identities. Developers who repeatedly
publish malicious code may face consequences such as loss of user trust or
banning from the repositories. For example, a package mirror which publishes
malicious code may be removed from future lists of mirrors, or a developer who
takes over a browser extension and publishes a malicious version~\cite
{extension-takeover-zdnet} may be blocked from publishing future code
updates.

Again, web applications fail to have accountability. A malicious or
compromised web server may publish malicious code to certain users, but
there is no public record of the code which it serves, and thus no way for
users to hold the server accountable. 


%

\noindent Summarising, it is difficult to establish trust in the web
 as a software distribution mechanism because it lacks \emph{auditability}
 (the means for anyone to inspect the code being distributed to others) and \emph
 {accountability} (the means to hold a developer accountable for the code
 they publish).

In this paper, we propose an opt-in transparency protocol that aims to
establish more rigorous trust relations between browsers and web
applications, and provide the foundation for 
a more secure web. Using our standard for accountable delivery of active content,
efficient and easy-to-use code-signing technique, and public transparency logs;
websites can convince the users that they are trustworthy in an economical way.
%
%
%
At a high level, we propose that web application developers, who choose to opt-in,
provide a signed manifest enumerating all the active content in their applications.

The
manifest files in our proposal are stored in publicly readable transparency logs. When a browser
requests a URL and downloads the resulting HTML document from the web
server, the web server also provides the corresponding manifest for this
URL. The browser checks that the active content provided by the server
matches the manifest entry, that the manifest is correctly signed,
and that the provided manifest is
consistent with the transparency logs.
    
Moreover, our proposal aims to reinforce the communication between the browser
and the web server by adding non-repudiation to the HTTP request-response
procedure. By itself, \ac{TLS} does not provide evidence that what was delivered
actually originated from the web server.
    %
Using digital signatures, we show how HTTP requests can be extended to provide a proof of origin.
%

From the signed manifest, the transparency logs, and the non-repudiation mechanism,
the protocol establishes that:
    \begin{itemize}[leftmargin=1.2em]
        \item The code a user executes is the same for
        \begin{nnew} the users of the plugin within a certain timeframe 
        \fixed{\reviewnote{R\ref{r1}.\ref{r1:same-for-everyone}}}
        depending on the validity of the manifest and a new manifest is signed.
        \end{nnew}
        \item On the client side, the code is bound to interact with
            third party code according to how the developer declared
            in the manifest. This includes the order of execution, the
            trust relation to third party code, and the use of
            sandboxing.
        \item If the code's execution is inconsistent with the manifest,
            the browser can provide a claim that can be verified by
            the public.
      \end{itemize}

\noindent Our proposal can be implemented by changes in the server
 configuration only, without the need to modify the served web content
 (assuming that the web page already makes use of \acl{SRI}
 hashes) and without changes to the HTML standard.



\medskip \noindent To sum up, our contributions are as follows:
\begin{enumerate}[leftmargin=1.3em]
    \item We propose Accountable JS, a protocol to enable auditability and
     accountability for web apps.
    \item We formally model Accountable JS with the \textsc{Tamarin} Prover and prove desired properties in the presence of active adversary.
    \item We implement Accountable JS in a browser extension that obtains the
     signed manifest, verifies its signature, and both statically and
     dynamically ensures that the active content on a web page agrees with
     the manifest. We also provide a code-signing mechanism for the
     developers.
     \item We evaluate the deployment of this technology
         and the performance overhead for the client in six case
         studies, including real-world applications: 
         Google AdSense, Nimiq and 
         WhatsApp.
    \item We model Meta's Code Verify protocol and compare its properties with Accountable JS.

\end{enumerate}

\paragraph{Relationship to Meta's Code Verify protocol}

In ~\cite{codeVerify}, Meta (formerly Facebook) proposed Code Verify,
likewise implementing a mechanism to enforce accountability via
transparency for active content in the web.
Our present proposal goes beyond Code Verify and provides a superset
of its functionality, most notably the ability to
delegate trust to third parties. On the other hand, our browser
extension is an academic prototype and thus not ready for productive
use. The protocol has the same message flow, but chooses a different
signature scheme and encodings.
We elaborate on these differences in \cref{sec:code-verify}.
%
%
An initial draft of the present proposal was shared with Meta's WhatsApp team
in 2022.
The protocol, manifest file format and browser extension
we present in this work are academic developments by the authors and
not endorsed by Meta in any way. 

%% file: background.tex
\section{Background}\label{sec:background}

Web pages are delivered via HTTP or HTTPS. In the latter case, a secure
and authenticated \ac{TLS} channel tunnels the HTTP protocol. 
\begin{nnew}
Typically,
the initiator of the \ac{TLS} connection, i.e. the web browser, is not
\fixed{\reviewnote{R\ref{r1}.\ref{r1:role-owner}}}
authenticated\footnote{At the communication layer. Authentication may
be implemented at the application layer.}, whereas the responder, i.e.
the web server, is identified with their public key and a certificate
linking the public key to the domain.
\end{nnew}

The authentication guarantees of \ac{TLS} exclude 
non-repudia\-tion of origin,
i.e. a communication party cannot prove
to a third party that they received a certain message. This property
is an important building block for accountability and can be achieved,
e.g. using digital signatures.
After the shared keys are established in \ac{TLS}, any messages exchanged 
could be produced by either party.  Roughly speaking, the party
providing the evidence has enough information to forge it.
Ritzdorf et al.~\cite{ritzdorfTLSNNonrepudiationTLS2018} proposed a \ac{TLS} extension that provides
non-repudiation, but it has
not been deployed in the wild.

Browsers typically parse the HTML document describing the web page
into a tree of HTML elements  called
\ac{DOM}\fcite{DOM-LEVEL-3-VAL}.
Some HTML elements have
\emph{active content}, which includes Flash or
Silverlight, but we will focus on \ac{JS} in this work. Active content can
be
\emph{inline},
i.e. hard-coded in \texttt{<script>}-tags or event handlers,
\emph{external},
i.e. referring to an external \ac{JS} file by URL,
or \emph{via iframe}, i.e.
the web page contains an iframe that refers to an HTML file which,
again, contains active content.
\begin{full}The browser may include multiple windows with multiple tabs,
displaying websites in parallel. For our purposes, we can abstract the
browser to a single top-level window that represents the client side
in the $\Http$ protocol.
App stores provide a unified distribution system for applications.
Typically, they are curated by their owners: developers submit their
software, the app store owner inspects them for compliance with
their guidelines (which can include quality control, but also
censorship) and distributes them to their users. While it has been
proposed to use app stores to deliver diversified versions of
software\fcite{franzUnibusPluramMassivescale2010} and both Google and
Apple's App Store support A/B testing, users expect to receive
untargeted applications.
\end{full}

Like in the case of app stores, we distinguish the roles of
the \emph{website}, which is distributing the web application, and the
\emph{developer}, which is the author of the web application.
This allows us to view the website as a distribution mechanism that is
necessarily online and publicly visible, as opposed to the developer,
who can be offline most of the time.
We distinguish the following roles:
\begin{itemize}[leftmargin=1.2em]
    \item The web application developer (short: \emph{developer}) creates the active content and has
        a secure connection to the web server. It is not active all the
        time.
    \item The web server (short: \emph{server}) delivers code provided by the developer to
        the \emph{client}. The website and the developer are
        associated with a domain, but the client is anonymous.
    \item The web browser (short: \emph{client}) requests a URL from the website.
\end{itemize}

A transparency log (short: \emph{ledger}) provides a publicly accessible database.
It typically has the property of being append-only (for consistency), auditable,
verifiable, and it hinders equivocation.
Hence, for the data in the logs, all parties are convinced that it is 
a public record and that everyone sees the same version of it.
We are using the ledger to store manifest files for each URL.
Having public records of the manifest files allows us to reason about accountability.

\subsection{Threat Model}\label{sec:threat-model}

\paragraph{Dolev-Yao attacker}

We consider a Dolev-Yao style adversary, i.e.
cryptography is assumed perfect (i.e. cryptographic operations do not leak 
any information unless their secret keys are exposed), but the
attacker has full control over the network.
This is formalised in our SAPiC~\cite{kremer2014automated} model in
\cref{sec:protocol-verification}.
Informally, we assume hash function to behave like random oracles,
signature schemes to be unforgeable and \ac{TLS} to implement an authentic
and confidential communication channel. We also rely on an intact
public-key infrastructure.

\paragraph{Corruption scenarios}

We assume honest parties follow the protocol specification and
dishonest parties are controlled by the attacker. The parties which considered honest  
are determined by the property of interest: 
\begin{itemize}[leftmargin=1.2em]
    \item  \emph{Accountability and Authentication of Origin:}
        An honest client wants to be sure that code is executed only if it 
        was made public and transparent i.e. inserted into logs by the developer; 
        here developer and web server are assumed dishonest.

\item \emph{Non-repudiation of Reception}
    A dishonest client may want to present false evidence for having
    received some \ac{JS} code. Here we assume the public is trusted and
    run a specified procedure\footnote{Detailed in Appendix
    ~\ref{sec:claimver}.} to check the evidence, and the web server to
    behave honestly, i.e. not to help the client provide false claims of
    reception, which are against the web server's interest. 

\item \emph{Accountability of Latest Version} 
    An honest client that receives a version of the code and wants to
    ensure it is the latest version. We assume an honest global clock
    that helps comparing the time of the code reception and the latest
    version at that time, and consider a dishonest developer and
    web server.
\end{itemize}

\paragraph{Target websites}
We target developers that \emph{aim at establishing user trust} or
pretend to do so.  Hence we assume, for honest developers, that active
content changes infrequently, e.g. multiple times per day, and that
their code facilitates the audit. Dishonest developers may
counteract, but, due to accountability and authentication of origin,
it is publicly recorded. 

Therefore, while our formal security arguments make no assumption on
how often the code changes are or how obfuscated it is,  we assume
that, from accountability of authentication of
origin, code obfuscation attacks or microtargeting are practically
disincentivised.

\paragraph{Browser features \& Transparency log } We assume 
the current browser security features, specifically the \texttt{sandbox}
attribute of the iframe tag, to be implemented correctly.
Furthermore, the
transparency log is trusted, efficient, available,
append-only and provides non-equivocation (i.e. the same information
is served to everyone). 
\begin{nnew}
Many strategies are available to implement such
a log. For example,
Trillian~\cite{trillian} and CONIKS~\cite{coniks190974}
use data structures that can be distributed over multiple parties and
allow to prove append operations efficiently.
\fixed{\reviewnote{R\ref{r3}.\ref{r3:transparency-log}}}
Misbehaviour can thus be detected by trusted public  auditors or by
honest logs distributing such proofs (called \emph{gossiping}).
See \cite{meiklejohnSoKSCTAuditing2022a} for a survey over different
mechanisms.
\end{nnew}

%% file: use-cases.tex
\section{~Use Cases}\label{sec:use-cases}\label{sec:types-of-webapp}

We introduce several types of web applications that will benefit from our protocol.
We will revisit these examples later and show how our approach can be applied to them.
%

\subsection{Self-Contained Application} Perhaps the simplest possible web
application is a one-page HTML document with active content 
that simply prints `Hello World' into the developer console. 
Upon loading this website, a user can manually check that its sole behaviour was
to print `Hello World', but they have no guarantees about subsequent page loads:
a server could easily decide to provide different behaviour to certain
users, or to insert malware based on IP address or browser fingerprint. For
this simple example, the consequences of a malicious or compromised server are
relatively limited, although we remark that cryptojacking\footnote{Malicious \ac{JS} which secretly mines cryptocurrencies in unsuspecting
users' browsers.} is a growing trend~\cite{DBLP:journals/ieeesp/CarlinBOS20}.
\begin{full}
We remark that, by default, every user who loads this web application receives the same
source code. However, there is no easy way for users to verify this fact.
\end{full}
%

%
\begin{full}
More complicated web applications may have login functionality, or
asynchronous client-server communication, or other advanced features. In
order to personalise users' experiences, applications may dynamically fetch
data using technologies such as Relay \footnote{\url{https://relay.dev/}} or Apollo
\footnote{\url{https://www.apollographql.com/}}. However, it is often still the case
that all users receive the same \ac{JS} source code bundle.
\end{full}

WhatsApp Web is a large real-world self-contained web application: its source code is
bundled using WebPack and served to all users; personalisation is implemented
through local storage and dynamic data fetching.
We will show how our protocol can be applied.

\subsection{Trusted Third-Party Code}\label{sec:ex-trusted-third-party}

Many websites rely not just on their own content but on resources served by a
third party. This may be a \ac{CDN} serving common
\ac{JS} libraries, embedded content such as photos or videos, analytics
and measurement libraries, tracking pixels, fraud detection libraries, or
many other options. For example, the following code loads the jQuery
\ac{JS} library from a \ac{CDN}, and uses it to display a `Hello World'
message.

\begin{lstlisting}[language=html,caption={Trusted third party code},captionpos=b, label={html:Trusted-third-party-code}]
<html><head>
  <script src="https://googleapis../jquery-3.6.1.min.js" integrity="sha384-i6..."/></head><body>
  <script>$("body").html("Hello World")</script>
</body></html>
\end{lstlisting}

As before, users are supposed to always receive the same code from the server. This time,
there is an additional avenue for compromise, though: even if the first-party
server is honest, it is possible for the \ac{CDN} to perform targeted attacks.
The developer, however, wants to pin the third party code to the
precise version that they inspected or trust.

\subsection{Delegate Trust to Third Parties}

The application uses third party code that its developer
cannot vouch for. This can be the case if the code is too complex to
inspect or if the application developer wants to always
use the latest version.
The third party developer, however, is willing to vouch for their code.
An example of this is Nimiq's Wallet, a web application for easy
payment with Nimiq's crypto currency. This application can be embedded
by first-party applications that provide, e.g. a web shop, who are
willing to trust Nimiq, but only given that they make themselves
accountable for the code they deliver.


\begin{lstlisting}[language=html,caption={Delegate trust to third party},captionpos=b, label={html:delegate-trusted-third-party-code}]
<html><body>
 <script type="text/javascript">
  function addTransaction () { 
    window.postMessage({'id': '123', 'amount': '10n', 'from':'abc'}, 'https://wallet.nimiq.com/');}
 </script>
 <iframe src="https://wallet.nimiq.com/" onload="addTransaction()"></iframe>
</body></html>
\end{lstlisting}

\subsection{Untrusted Third-Party
Code}\label{sec:untrusted-third-party}

For web technologies, consecutive deployability is a must. Hence, in
this use case, the application developer cannot audit the code, 
but the third party does not use Accountable JS. The
application developer needs to blindly trust the third party, but
using sandboxing techniques, it can restrict the access that the
possibly malicious script provided by the third party can have.

A particularly important instance of this problem is ad bidding. The
third party is an ad provider that decides online which ad is actually
served. Because they cannot review the ads that they distribute, 
which may contain active content, they are not willing to vouch for
the code they distribute. 
This is the case for Google AdSense, used by over 38.3 million
websites.
Cases where ads were misused to
distribute malicious code are well documented~\cite{Adsense-policy}.

\subsection{Code Compartmentalisation}

The application that the developer provides can be compartmentalised
so that the most sensitive information is guarded by a component that
is easy to review and changes rarely.
The other components that are user-facing and changing more often are
separated from this component using sandboxing. 
The developer wishes to reflect this structure and make themselves
accountable for the whole code, but also separately commit on keeping
the secure core component small and auditable.
For example, Nimiq's Wallet components follow a similar structure.



%% file: approach.tex
\section{~Approach: Accountable JS}\label{sec:approach}


We propose a cryptographic protocol between 
the client,
the server,
the developer, 
and
a distributed network of
public transparency logs.
The protocol's objective is to hold the developer accountable
for the code executed by the browser.
The protocol 
provides four main functionalities:
\begin{itemize}[leftmargin=1.2em]
    \item The server provides a manifest declaring the active content
        and trust relationships of the web application, which the
        client compares with a published version on the transparency
        logs.
    \item The client measures and compares the active content received
        by collecting active elements, 
        e.g. \ac{JS}, in the HTML document delivered by the web server.
    \item Developers and clients submit manifests to a public
        append-only log to verify that everybody receives the same
        active content.
    \item The server signs a nonce as non-repudiable proof of origin
        for the \ac{JS} that the client receives.
\end{itemize}

\paragraph{Website Manifests}

Website developers may provide a signed manifest for each publicly 
    accessible URL in their website (excluding the query string).
    The signed manifest comprises a manifest and a signature block over it.
    A manifest describes the webpage, including, besides the active
    content, its URL and a version number.
    The active content is described in a custom format. 
    We elaborate on the
    manifest directives in the supplementary material ~\cite{supplementary-files}.
    The developer's identity is distinct from the server's\begin{nnew}, 
    but their certificates must share the same Common Name(CN) to 
    \fixed{\reviewnote{R\ref{r3}.\ref{r3:validity-terms}}}
    restrain from unauthorised manifest deployments.\end{nnew}
    The
    browser validates 
    the authenticity of the developer's public key 
    in the same way, using
    the existing \ac{PKI} and 
    its built-in root \ac{CA}
    certificates.

Accountable JS is an opt-in mechanism. The website declares the signed
manifest using an experimental HTTP response header field 
called \accjsheader.
Henceforth, the client, however, expects the website to provide
a valid manifest for this URL in any case. 

\paragraph{Client Measurement}

The client measures the active content inside the HTML document
delivered in the response body, collecting information about each
active element in the document and validating it with the
corresponding manifest block in a manifest file.
Elements that cannot be matched trigger an error and the user is
warned about this error. The current extension is not preventive, but
in the future with pervasive developer support, browsers may choose to
halt the execution if delivered code is inconsistent with the
boundaries drawn by manifest.
%
%
The active content is measured with a so-called \emph{mutation observer}, starting with the first request. The measurement procedure that we developed listens to the observer’s collected mutations that regard active elements in a list.
In Section~\ref{sec:measurement}, we explain the process in more detail.

\paragraph{Manifest Logs}

While a signed manifest may prove the integrity and authenticity of the manifest,
    it cannot prevent equivocation, 
    i.e. it cannot prove the same signed manifest is delivered 
    to every request by the web server. 
    To this end, we propose to use transparency logs.
    A manifest file declares a version number and the version number is unique
    per manifest file.
    The developer publishes their signed manifest in a publicly
    accessible, auditable, append-only log
    like the 
    \ac{CT} protocol~\cite{rfc6962}, 
    which provides logs for \ac{TLS} certificates. 
    Clients may verify that a version they receive is the latest
    online, or use a mechanism like \ac{OCSP}-Stapling~\cite{rfc6961} to check that
    a version they receive was the latest version a short time
    ago. 
    Any client that encounters a signed manifest that is not yet in the log
    can submit it to the log.
    We discuss the transparency log considerations in more detail in
    \cref{sec:transparency-logs}.

\paragraph{Non-Repudiation of Origin}
\begin{conf}
We propose a simple non-repudiation mechanism for the client's web requests, 
so that in case a developer distributes damaging active content, a client can prove that they have received that content from a web server.
The client transmits a nonce via a request header and the server signs this nonce
along with the signed manifest (c.f. Section~\ref{sec:protocol}).
\end{conf}
\begin{full}
We propose a non-repudiation mechanism for the client's web requests.
In case a developer distributes damaging active content, a
client cannot prove that they have received that content from
a server.
While \ac{TLS} provides integrity of communication via Message Authentication
Codes and authenticity of the communication partner via its handshake,
the client is nevertheless unable to prove that they received damaging
content 
as both communication partners can forge the message
transcripts after the key exchange. 

We propose a simple mechanism whereby the server signs a nonce
chosen by the client, along with the signed manifest. 
The client transmits this nonce via a request header.
We elaborate on the non-repudiable web request protocol in
Section~\ref{sec:protocol}.
\end{full}



%% file: manifest-intro.tex
\section{Manifest File}\label{sec:manifests}

In the manifest, the developer declares the active elements a web
application
is bound to execute during its run time. The run
time starts from the web request and ends with the window's close or a new
web request. For \ac{SPA} (e.g. Nimiq), the run time for the web page ends when page is refreshed, its URL is changed 
or the window is closed.

The manifest file represents the active elements and their relevant
metadata as a collection of attribute-value pairs in the JSON format.
The metadata expresses the trust relations w.r.t.\ third party content
and settings for sandboxing. 
	%
        The top-level properties in the manifest, also called manifest
        header, contain descriptive information about the web page:
	its URL,
	its version number,
        and optional metadata, e.g. the developer's email address. 
        The domain within the URL determines which keys can be used
        to sign the manifest, namely, the common name of the
        signature key's certificate has to match that
        domain.\footnote{%
	The query component of the URL\fcite{RFC3986} can be excluded, 
        since the browser extension discards that part in the measurement.}
	The developer can decide for any numbering scheme for the
        version,
        but they must be strictly increasing with each new manifest
        published.

A manifest file is accepted if it is
\emph{syntactically correct}, 
i.e.\ follows the schema (see manifest manual in the supplementary material ~\cite{supplementary-files}
for details), 
\emph{complete}, i.e.\ it contains enough information about 
the web application and its active elements to enable evaluation,
and, most importantly, \emph{consistent with the delivered resource},
i.e.\ that evaluation succeeds.

%% file: manifest-order.tex
\subsection{Execution Order}\label{sec:exec-order}


	An active content is considered dynamic 
	if it is added after the window's load event; otherwise, it is static. The manifest specifies 
        elements as either static or dynamic using the
        $\dynamic$ attribute.
        \acp{SPA} in particular download or preload
        resources during navigation, rewriting the \ac{DOM} on the fly
        depending on how the user navigates.

        \begin{nnew}
        For static elements, the sequence number $\manseq$ specifies in
        which order they must appear after browser renders the
        delivered HTML. 
        \fixed{\reviewnote{R\ref{r2}.\ref{r2:seq-meaning}}}
        It starts from $0$ and repetitions are not allowed.
        Dynamic content is only measured if they are present in the web page, i.e. 
        it is allowed to be injected, but not required to.
        This mechanism can also be used to declare region-specific active
        content. 
        The order is ignored for dynamic content.
        \end{nnew}
	The measurement procedure will check if the list of the elements in the manifest
	is in the same order except 
	for elements that will be dynamically added to the \ac{DOM}. 
        Elements may be removed dynamically, but
        only if the attribute
        \persistent is set to false.

\begin{full}
        A \ac{JS} element can be loaded synchronously ($\sync$), asynchronously ($\async$) or it can be deferred until the HTML parsing is done ($\defer$).
    A synchronous  \ac{JS} element blocks the HTML parsing process and
    is executed in-order.
    Asynchronous and deferred elements do not block parsing.
    Asynchronous elements are loaded in parallel with other HTML
    elements, while deferred elements are loaded after parsing has
    finished. 
     Hence, for both, the position on the \ac{DOM} tree may not be predicted precisely. 
    %
    %
\end{full}

%% file: manifest-trust.tex
\subsection{Trust and Delegation}\label{sec:manifest-trust}

With the manifest, the developer provides assurance for the active
content in their application.
    Third-party components, e.g. 
    \ac{JS} libraries, bootstrappers, advertisements or ad-analytics tools
    play a significant role in most modern web applications, which are
    thus a
    mixture of first-party code and code from multiple third parties.
    In the manifest, we enable the developers to decide 
    the trust level on each active element imported to their web applications.
    For instance, they can take the responsibility and provide assurance 
    (i.e. with a cryptographic hash) on first party elements 
    while for the external elements, they may declare a valid source and delegate 
    the trust on the developers of those resources. 
    
We thus require each block in the manifest to have a trust declaration.
    There are three options to declare the trust level:
    
    \begin{itemize}[leftmargin=1.2em]
        \item $\assert$ : The developer provides the hash of the
            expected active content and asserts it is behaving as
            intended. It is computed using the standard \acl{SRI} hash generation
            method~\cite{Using-SRI}, i.e. comprises the hash
            algorithm used, followed by a dash and the base64-encoded
            hash value.

        \item $\delegate$: 
            The developer refers the trust to the third party
            providing this element. Now the third party is taking
            responsibility for this code and provides a manifest whose
            location is either declared in the first-party manifest,
            or delivered in the headers of the third party's response.
            The third party manifest can likewise delegate trust,
            thereby constructing a chain of trust delegations.

        %
        \item $\blindtrust$: 
            The developer blindly trusts the third party, without
            identifying the code they trust.
            This should only be used with the
            $\mansandbox$ attribute.
    \end{itemize}

%% file: manifest-types.tex
\subsection{Types of Active Elements}\label{sec:write-manifest}

\input{trust-table}

	The developer describes the manifest blocks for each active element 
	by their resource type \mantype
	(e.g. $\mathjavascript$, $\iframe$),
	trust policy $\mantrust$
	(e.g. $\assert$, $\delegate$, $\blindtrust$),
        whether they are dynamic or static and,
        in case they are static, their sequence number
        $\manseq$.
	%
        %
        There are mandatory and optional directives for writing a manifest and these directives may depend on the resource type.
        %
        %
        If the developer declared a manifest section including an optional directive, that does not mean this directive is ignored in the evaluation; this directive still is part of the evaluation.
        For instance, the $\mancrossorigin$ directive is optional for $\external$ resource type, but if the developer declares a $\mancrossorigin$ attribute, then it has to match with the active content information.
        Not all resource types support all trust policies (see
        Table~\ref{tab:trusttypes}). We will discuss them one by one:
	\begin{itemize}[leftmargin=1.2em]
            \item $\inline$: Inline scripts are \texttt{script}
                elements without the
                     $\mansrc$ 
			attribute, i.e. the \ac{JS} code is included in
                        the
			HTML document.
			Therefore, $\mantrust$ can only be $\assert$
                        and may be omitted.
                        The cryptographic hash covers the included \ac{JS}
                        code, i.e. the textContent value of the script element.
			
		\item $\eventhandler$: Event handlers are active
                    content included in attributes such as
                    \texttt{onClick} that are executed on HTML
                    events. Like inline scripts, \mantrust must be
                    \assert and can be omitted.
                    \begin{full}
			Unlike $\inline$, however,
                        the (\ac{SRI}-encoded) hash value covers the entire element, including 
			the HTML tag itself.
                    \end{full}

                    \item $\external$ : A \texttt{script} element can
                        be outsourced by specifying its URL in the
                        \mansrc attribute.
			%
			%
			An $\external$ script can originate from 
			a different origin (cross-origin) or from the same origin.
			%
                        Trust can be set to $\assert$ and
                        $\delegate$ -- as sandboxing is not supported
                        for external scripts, $\blindtrust$ would give
                        little assurance. 
            %


                        %
		\item $\iframe$ : An $\iframe$ embeds another document within the current document.
                        There are three ways this can
                        happen, which the manifest file represents
                        using the attribute $\mansrctype$.
                        The most common is to specify a URL
                        (\mansrctype = \external). 
                        \begin{conf}
                        The other ways 
                        ($\mansrctype = \mansrcdoc$ and \mansrctype = \script) are explained in the full version~\cite{full-version}.
                        \end{conf}
                        This type of content can be declared with 
                        any $\mantrust$ value.
                        \begin{full}
                        For $\mantrust = \assert$, it is possible to
                        either hash the whole embedded HTML
                        document, or to provide a list of manifest
                        blocks for the active elements inside the
                        embedded document.
                        
                        Second, an iframe's content can also be hardcoded in
                        the outside document ($\mansrctype
                        = \mansrcdoc$), in which case \mantrust can only
                        be set to \assert (and the hash is computed on
                        the \mansrcdoc attribute of the \iframe
                            element).

                        Third, the embedded document can be created
                        using the $\documentwrite$ 
                        method\fcite{Document-write}, via \ac{JS} code
                        inlined
                        in the iframe's \mansrc attribute (\mansrctype
                        = \script).
                        Since this \ac{JS} code is known at this point, \mantrust
                        must be set to \assert and, again, either the
                        \ac{JS} code's hash be provided, or a list of
                        manifest blocks. 
                        \end{full}

                \end{itemize}

%% file: trust-table.tex
\begin{table}
        \renewcommand{\cmark}{$\bullet$}
        \renewcommand{\xmark}{$\circ$}
        \centering
        \caption{Trust Relationships by Type of Active Element\label{tab:trusttypes}}
        \begin{tabular}{lcccc}
        \toprule
        &  \multicolumn{3}{c}{\mantrust} & \\
        \cmidrule(r){2-4}
        \mantype  &  \assert & \blindtrust & \delegate & \mansandbox \\
        \midrule
         \inline & \cmark & \xmark & \xmark & \xmark \\
         \eventhandler & \cmark & \xmark & \xmark & \xmark \\
         \external & \cmark & \cmark & \cmark & \xmark \\
         \iframe with \ldots \\
         ~~\mansrctype=\external & \cmark & \cmark & \cmark & \cmark \\
         ~~\mansrctype=\mansrcdoc & \cmark & \xmark & \xmark & \cmark \\
         ~~\mansrctype=\script & \cmark & \xmark & \xmark & \cmark \\
         \bottomrule
        \end{tabular}
        \vspace*{-4mm}
\end{table}

%% file: manifest-sandboxing.tex
\subsection{Sandboxing}{\label{sec:man-sandboxing}

Besides, iframes permit the use of sandboxing via the attribute
with the same name~\cite{Eicholz:21:H}.
A sandboxed iframe is considered a cross-origin resource, 
even if its URL points to the same-origin website.
Hence, because of the browser's same-origin-policy\fcite{RFC6454}, 
the parent window and the iframe are isolated, 
and they cannot access the \ac{DOM} of each other.
Furthermore, sandboxing blocks the execution of \ac{JS} and the submission
of forms and more. These restrictions can, however, be lifted
using an allow list in the HTML tag.

As we will see in the next section, security-critical websites need to
use sandboxing to protect data from other browsing contexts; hence
we reflect the $\mansandbox$ feature in the manifest file. 
The measurement procedure ensures that the active
element has an equally strict or stricter sandboxing policy than
described in the manifest. 
\begin{nnew}
\fixed{\reviewnote{R\ref{r3}.\ref{r3:sandboxing}}}
An allow list is stricter if it is a subset of the other.
\end{nnew}

%% file: case-studies.tex
\section{~Use Cases, Revisited}\label{sec:case-studies}

We come back to the use cases from
Section~\ref{sec:types-of-webapp}
to illustrate how
Accountable JS applies to real-world web applications with
different trust assumptions.
%

\subsection{`Hello World' Application}

We begin with the basic `Hello World' website example, and add
a reference to the manifest in its meta tags. 

\begin{lstlisting}[language=html,caption={First example: Hello World.},captionpos=b, label={html:deliver-manifest}]
<html><head>
  <@\textcolor{red}{\bf\verb~<meta charset="utf-8" name="x-acc-js-link" content="http://www.helloworld.com/manifest.sxg">~}@>
</head><body>
  <script>console.log("Hello World")</script>
</body></html>
\end{lstlisting}

Alternatively, the manifest can be provided as an HTTP response
header. The manifest file provides the URL and version of the website
and lists the base64-encoded SHA-256 hash of the
inline script. 
\begin{lstlisting}[language=json,caption={Manifest for first example.},captionpos=b, label={manifest:simple-example}]
{ "url": "http://www.helloworld.com/",
  "manifest_version": "v0",
  "contents": [
    { "seq": 0,
      "type": "inline",
      "load": "sync",
      "trust": "assert",
      "hash": "sha256-AfuyZ600rk..."}]}
\end{lstlisting}

\subsection{Self-Contained Web Applications}\label{sec:serverless}

Web applications can be completely self-contained. This may be for security
or because they follow the recent serverless computing
paradigm (e.g. Amazon~Lambda).
In serverless computing,
a web application developer may only write static user-side code and
delegate all the server-side logic to a cloud service provider.

The application of Accountable JS is straightforward in this case:
as part of our prototype, we developed our deployment tool 
$\generateManifest$,
which computes the hash values of all active contents in the browser and
produces a manifest file that asserts their trustworthiness. The
developer can then sign this manifest file.

We tested this methodology on a popular example, the WhatsApp Web
client, and provide the manifest file 
in the 
supplementary material~\cite{supplementary-files}.
It lists nine external and four inline scripts.


\subsection{Trusted Third-Party Code}

The developer can use the manifest file to identify the included
third party code by hash and set the order of execution.
This expresses that the developer vouches for the third party code.
We add the following attribute to the header of
`Hello World' example from \cref{sec:ex-trusted-third-party} and we declare it in the manifest file with $\mantrust=\assert$.
\begin{lstlisting}[language=html]
  <script src="https://googleapis../jquery-3.6.1.min.js" integrity="sha384-i6..."></script>
\end{lstlisting}



\subsection{Delegate Trust to Third Parties}

The first party can delegate trust to a third party by embedding their
code in an iframe (or linking their \ac{JS}) and setting 
\mantrust to \delegate.
The extension will verify the
third party code based on a manifest file signed by its developer.
This expresses that the main developer vouches for the third party to
be trustworthy, but demands that the third party itself can be held to
account. This is in contrast with trusting a concrete piece of code
provided by the third party.

We tested this technique using Nimiq's Wallet, which can be embedded
in third party web pages. These can now combine the code that they
control (e.g. for setting up a shopping cart) with the code that
Nimiq provides for signing transactions.

The website's manifest below (Listing~\ref{manifest:delegate-trust-third-party})
specifies some inline scripts with $\mantrust=\assert$ (omitted) and an
iframe with $\mantrust=\delegate$. 
The browser now expects the response to the query for the iframe's
content (\url{https://wallet.nimiq.com}) to point to a URL with a signed
manifest. 

\begin{lstlisting}[language=json,caption={Manifest is delegated to a trusted third party},captionpos=b, label={manifest:delegate-trust-third-party}]
{ "url": "https://www.example-shop.com/",
  "manifest_version": "v2",
  "contents": [
    [inline script manifests omitted]
    { "seq": 2,
      "type": "iframe",
      "src_type": "link",
      "src": "https://wallet.nimiq.com/",
      "sandbox": "allow-scripts",
      "dynamic": false,
      "trust": "delegate" }]}
\end{lstlisting}


\subsection{Untrusted Third-Party Code}\label{sec:adsense-use-case}

High-security applications may want to rely on third party code they
cannot vouch for, e.g. when including ads that are dynamically
chosen by an ad-bidding process. 
We developed a small web application that uses Google
AdSense and sandboxed this code, but noticed that AdSense and many
other ad providers require access to the top-level window~\cite{Adsense-iframe}
for fraud detection, e.g. to detect invalid
clicks\fcite{Adsense-iframe-prohibit}.

We therefore needed to turn the relationship between the secure code and the
untrusted code around. We sandboxed the secure code with \mantrust set to \assert,
protecting it from the potentially unsecure AdSense code, which is not
sandboxed and declared \blindtrust.
Now the AdSense code cannot access the secure document in the iframe. 
The manifest file is shown in List.~\ref{manifest:untrusteddelegate-manifest-nimiq}.
It includes thirteen active elements (six $\external$, seven $\iframe$) related to
    AdSense, along with
    Nimiq's Wallet (seq='6'), for which trust is delegated.

\begin{lstlisting}[language=json,caption={Untrusted AdSense and the Delegated Nimiq wallet at manifest section sequence number `6'.},captionpos=b, label={manifest:untrusteddelegate-manifest-nimiq}]
{"url": "https://www.helloworld.com/",
  "manifest_version": "v3",
  "contents": [
    [six external scripts for AdSense with trust=blindtrust]
    { "seq": 6,
      "type": "iframe", 
      "src_type": "link",
      "src": "https://wallet.nimiq.com/",
      "sandbox": "allow-same-origin allow-scripts",
      "dynamic": false,
      "trust": "delegate" // See Listing <@\ref{manifest:delegate-manifest-nimiq}@>
      },
    [six more iframes for AdSense with blindtrust]]}
\end{lstlisting}

\begin{full}
Note that we relax the sandbox attribute of the secure iframe to allow script execution 
and to gain access to its own origin (\url{https://wallet.nimiq.com})
to access its cookies.
Because of the trust delegation, we create
a new signed manifest, which we expand upon in the next section (see \cref{manifest:delegate-manifest-nimiq}).
The communication between the parts of the web application
that handle user interaction and the trusted code in the sandbox takes
place via postMessage calls.
\end{full}

\subsection{Compartmentalisation of Code and Development
process}\label{sec:case-compartment}

\begin{figure}
\centering
\includegraphics[width=\linewidth]{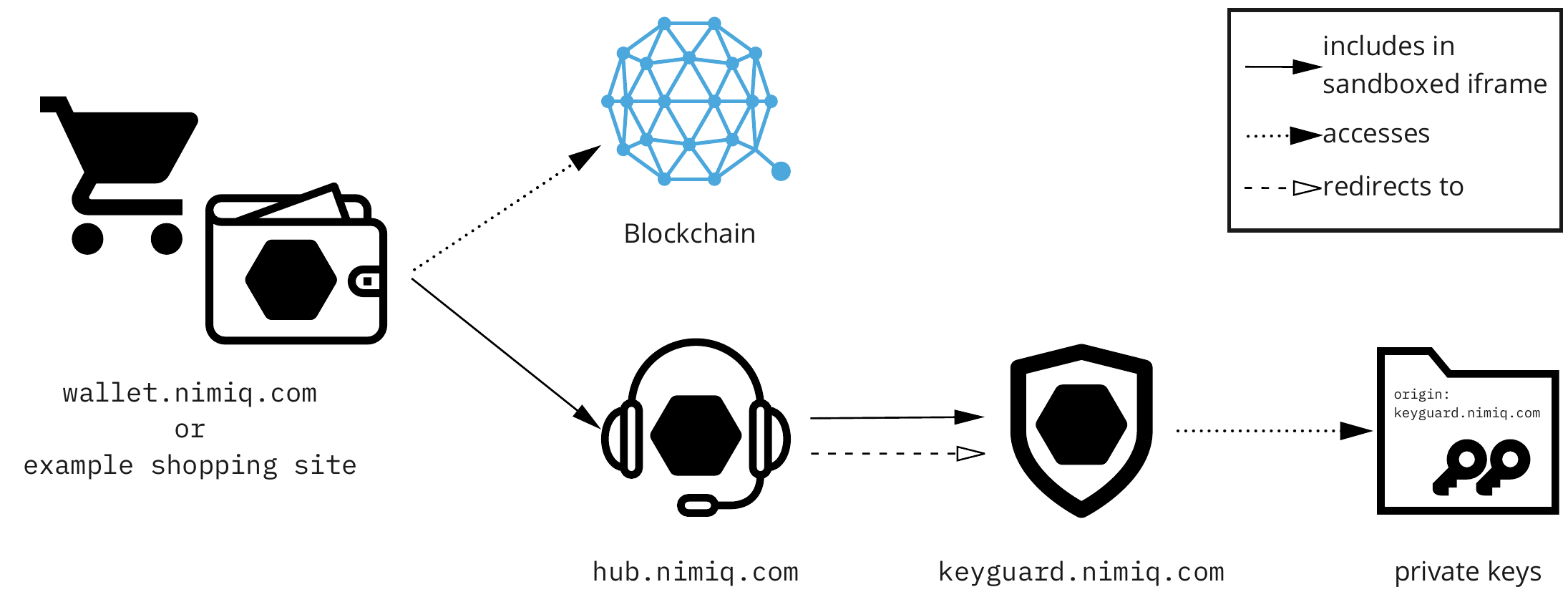}
\caption{Structure of Nimiq Ecosystem.}
\label{fig:structure-of-nimiq-s-wallet-application-}
\vspace*{-5mm}
\end{figure} 

We further expand on Nimiq's Wallet application, this time as an
example for compartmentalising the code and the signing process.
Nimiq's Wallet application at no point has direct access to the users' private keys.
It is treated the same way as any other third party application interacting with the Nimiq ecosystem
(see Fig~\ref{fig:structure-of-nimiq-s-wallet-application-}).
\begin{nnew}
It embeds 
the \emph{Hub} which acts as an interface to the user's addresses and can trigger actions on the private keys.
Access to the users' private keys is only possible through the Hub and pre-specified APIs.
\end{nnew}
The Hub will then forward any request that needs to access the private keys to the \emph{KeyGuard} component,
which upon user input can decrypt the locally stored keys, perform the requested action, and return the result to the Hub.

\begin{nnew}
The procedure $\generateManifest$ produces the following manifest 
for Nimiq's Wallet. 
Observe that it
heavily employs sandboxing. 
\end{nnew}
Both included iframes have the
\mansandbox attribute set empty, meaning no exceptions defined.

\begin{lstlisting}[language=json,caption={Delegated content Nimiq Wallet's manifest.},captionpos=b, label={manifest:delegate-manifest-nimiq}]
{"url": "https://wallet.nimiq.com/",
  "manifest_version": "v0",
  "contents": [
    [five external scripts]
    { "seq": 3,
      "type": "iframe",
      "src_type": "link",
      "src": "https://hub.nimiq.com/iframe.html",
      "sandbox": "",
      "dynamic": true,
      "trust": "assert",
      "manifest": [[seven external scripts],
        { "seq": 7,
          "type": "iframe",
          "src_type": "link",
          "src": "https://keyguard.nimiq.com/",
          "sandbox": "",
          "dynamic": true,
          "trust": "delegate"}]}]}
\end{lstlisting}

The Wallet's manifest includes \url{hub.nimiq.com} in an iframe,
containing, among other elements, the KeyGuard, which has a separate
origin and thus exclusive access to the user's keys.
For transactions, the Hub redirects to the KeyGuard.
The KeyGuard is trusted, easy to audit, does not depend on any
third party code and changes rarely. 
The KeyGuard manifest is as follows.
\begin{lstlisting}[language=json,caption={Nimiq Keyguard depends on its own content.},captionpos=b, label={manifest:keyguard-nimiq}]
{"url": "https://keyguard.nimiq.com/",
  "manifest_version": "v0",
  "contents": [
    { "seq": 0,
      "type": "external",
      "link": "https://keyguard../web-offline.js",
      "hash": "sha256-L8NMxOGkIW...",
      "load": "defer",
      "dynamic": false,
      "trust": "assert" 
    },
    [two external scripts w/ same dynamic/trust.]]}
\end{lstlisting}

The Wallet manifest file reflects the web applications compartmentalisation: 
every component -- Wallet, Hub and KeyGuard -- runs on a different domain, hence locally
stored information like the wallet key is inaccessible to the Hub or Wallet due to the same-origin policy. 

With this setup, it is easy to compartmentalise the development
process, too. A separate developer key could be used for the KeyGuard
code given that it is already bound to a second domain. New KeyGuard
releases would need to be signed by that key, which, internally, can
be assigned additional oversight requirements.
Without requesting a new key from the \ac{PKI},
any bypassing of this procedure would
either
end up with code that cannot access the user's key 
or
be provable with the signed manifest for the Wallet.





%% file: measurement.tex
\section{~~~Measurement procedure}\label{sec:measurement}



\begin{figure}
\centering
\includegraphics[width=\linewidth]{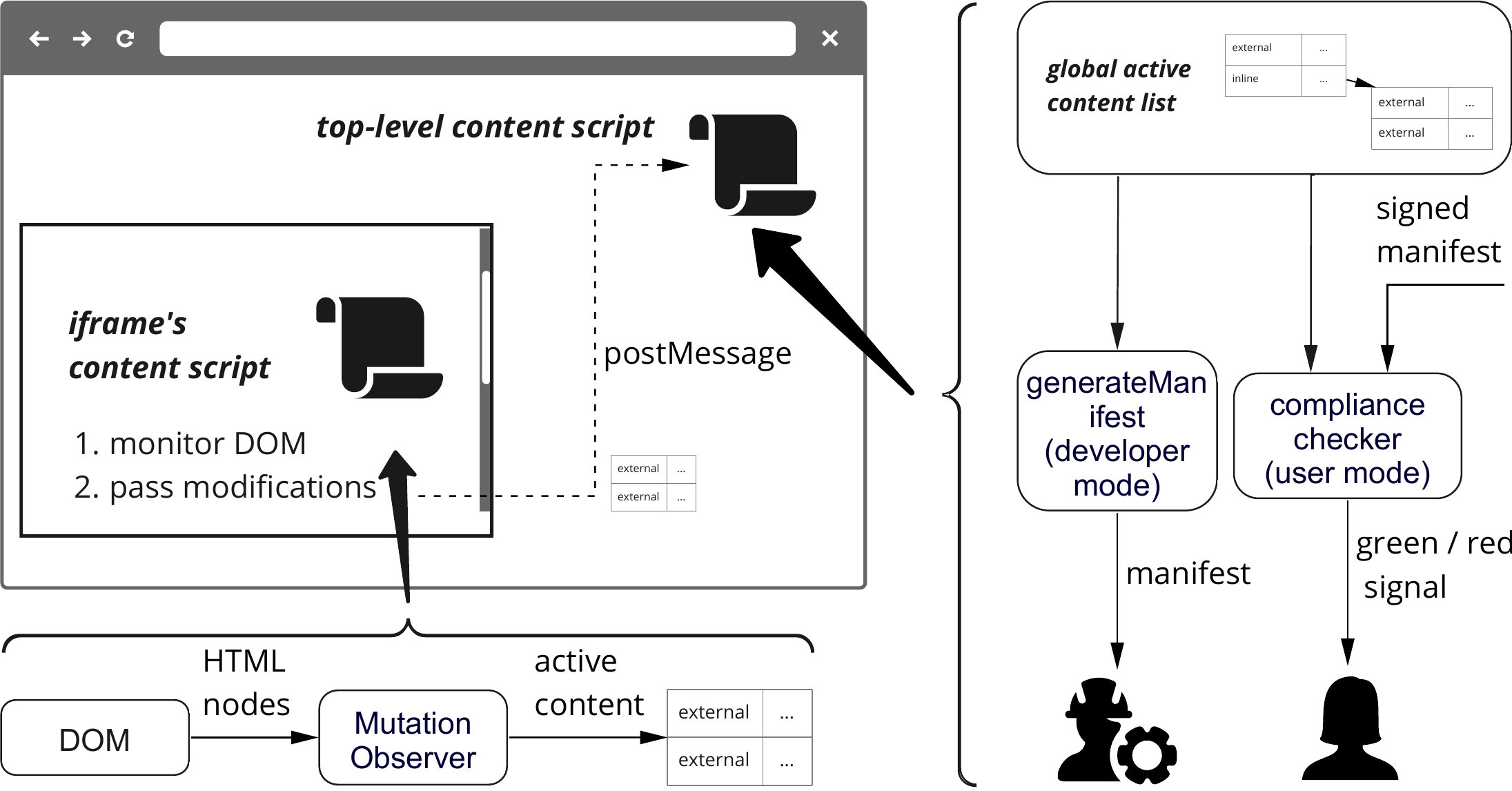}
\caption{Manifest file generation and metadata collection \label{fig:measurement-procedure}.}
\vspace*{-5mm}
\end{figure}

We present a practical active content measurement procedure
	that can be used to identify active elements and collect their metadata,
        allowing the client to check whether the web application complies
        follows
        the provided manifest.
        In development mode, the same procedure 
	can be used to automatically generate a manifest file 
	from an HTML document.
	
        The measurement procedure is depicted in Fig.~\ref{fig:measurement-procedure}.
	The browser's rendering engine parses the raw HTML 
	document and creates the \ac{DOM}, 
	observing the \ac{DOM} for mutations, e.g. elements that are
        added at run time.
        Whenever
	an active element
	is appended, edited or removed from the \ac{DOM},
	the metadata agent will be triggered,
	which keeps a list of the active elements and their metadata. 
	%

    The extension obtains access to the \ac{DOM} by defining a
    \emph{content script}, a script that runs in the context of the
    current page. This includes all pages loaded in top-level browser
    windows (e.g. Tabs), but also iframes within those.
    Content scripts running at the top level are responsible for
    collecting metadata on all active elements in their context.
    For nested iframes, they can only collect the metadata \emph{about} the iframe 
    like the attributes $\mansrctype$, $\mansrc$ and $\mansandbox$, but not inspect the document \emph{inside} this iframe.
    The same-origin policy forbids this in many cases.
    We therefore use the iframe's content script to gather information:
    if the content script recognises that it is not at the top-level, it
    runs statelessly, collecting the metadata as usual, but reporting it to
    the parent window's content script \begin{nnew} via postMessage\end{nnew}.

    \begin{nnew}
    The metadata agent distinguishes script and iframe elements by their HTML tags. 
    A script that has $\mansrc$ attribute is $\external$ otherwise it is $\inline$.
    For external scripts \ac{SRI} hashes, crossorigin and load attributes are collected.
    \fixed{\reviewnote{R\ref{r1}.\ref{r1:extension}}}
    For inline scripts, hash is computed on the script and the load attribute is collected.
    Event handlers are searched inside all \ac{DOM} elements checking if their attributes contain
    any of the global event attributes e.g. onclick in a list~\cite{event-handlers-html-spec}.
    \fixed{\reviewnote{R\ref{r4}.\ref{r4:event-handlers}}}  
    For event handlers, the hash is computed on the value of the event attribute.
    \begin{conf}
    For iframes with $\mansrctype=\external$, 
    the metadata agent in the parent window collects the crossorigin and sandbox 
    attributes and gathers the metadata about the document inside the iframe
    from its content script.
    \end{conf}
    \begin{full}
    For iframes, the metadata is collected based on $\mansrctype$ which is $\mansrcdoc$
    if the iframe has $\mansrcdoc$ attribute, otherwise $\script$ if the
    $\mansrc$ attribute has a script as a value,
    and $\external$ if the $\mansrc$ attribute has a URL value.
    For iframes with $\mansrcdoc$ or $\script$, 
    a hash is computed on the $\mansrcdoc$ or $\mansrc$ contents, 
    and crossorigin and sandbox attributes are collected by the metadata agent in the parent window.
    For iframes with $\external$, the metadata agent in the parent window collects the crossorigin and sandbox 
    attributes and gathers the metadata about the document inside the iframe
    from its content script.
    \end{full}
    Also, for each active element boolean $\dynamic$ and $\persistent$ scores are assigned by the metadata agent.
    An active content is considered dynamic 
    if it is added after the window's load event; otherwise, it is static.
    Elements that get removed from the 
        \ac{DOM} are marked to be non-persistent, 
        but still kept in the active content list 
    for evaluation.
    \end{nnew}

    \begin{full}
    An opt-in website could be opened in a popup or an iframe, the measurement procedure runs as usual in this case.
    If the opener/parent window is not opt-in, the measurement will only take the popup/iframe website into account.
    However, if the opener/parent window that is in the same-origin, it can
    cause changes in the popup/iframe context without triggering the mutation observer in the popup/iframe.
    This could undermine accountability, hence, we require the opener/parent window must also be opt-in if the popup/iframe is in the same-origin.
    \end{full}

	If the web page opted in, i.e. it has sent the
        \texttt{x-acc-js-link} header in the past and provided
        a valid manifest,
	then the metadata collector compares
        the metadata list
        with the list of active elements in the manifest.
	If the web page violates the protocol, 
        the extension reports this to the user.

        In developer mode, a failure to comply triggers
	the manifest generator to collect and 
	generate metadata for the active elements.
        The $\generateManifest$ procedure then produces a manifest
        file with $\mantrust = \assert$ for each active element
        based on the collected information, which can be easily
        adapted to other trust settings.
        This manifest represents the most restrictive manifest
        functional for this web application.



%% file: sxg.tex
\section{~~~~Signing and Delivering a Manifest}\label{sec:sxg}

A valid signature on the manifest proves that 
the manifest was created by a known origin, 
i.e. a developer publicly associated with the website,
and that it was not tampered with in transit.
To sign manifests, we adopt the \ac{SXG} standard\fcite{SXG-Webpackage}.
\ac{SXG} is an emerging technology that makes websites portable.
    With \ac{SXG}, a website can be served from others, by default
    untrusted, intermediaries (e.g. a \ac{CDN} or a cache server),
    whereas the browser can still assure that its content was not
    tampered with and it originated from the website that the
    client requested.
    \begin{nnew}
        This allows decoupling the web developer from the web host and 
        \fixed{\reviewnote{R\ref{r3}.\ref{r3:sxg-more}}}
        nicely fits our view of websites as software distribution
        mechanisms.
    \end{nnew}
    The \ac{SXG} scheme allows signing this exchange with an X.509 certificate 
    that is basically a \ac{TLS} certificate with the `CanSignHttpExchanges' extension.
    %
    Browsers will reject certificates with this extension if they are
    used in a \ac{TLS} exchange, ensuring key separation. 
    \ac{SXG} certificates are validated using the \ac{PKI}, allowing
    Accountable JS to be used
    with the existing infrastructure, although, currently, Digicert is
    the only \ac{CA} that provides \ac{SXG}
    certificates\fcite{Digicert-SXG}.
    The lifespan of an \ac{SXG} certificate is at most 90
    days\fcite{SXG-Webpackage}, limiting the impact of key leaks.

An \ac{SXG} signature includes the HTTP request, as well as the
corresponding response headers and body from the server.
The signature is thus bound 
to the requested URL, in our case, the manifest file's URL. 
It also includes signature validation parameters like the start and
end of the validity period and the certificate URL. 
If the current time is outside the validity period, \ac{SXG} permits
fetching a new signature from a URL. This URL is also contained in the
(old) signature's validation parameters.
These features provide a solid foundation for Accountable JS's signed
manifests, allowing manifests to be cached during the validity period
and enabling dynamic re-fetching and safe key renewals. 

A web application in compliance with Accountable JS must deliver the
signed manifest. If it is small enough, it can be transmitted directly
via the HTTP response header (using the directive $\accjsman$).
Alternatively, the response includes the URL of the  \ac{SXG} file,
using the HTML meta-tag or HTTP-response header $\accjsheader$.
The signature in this file includes
the manifest file (as the HTTP response body)
and the manifest URL (part of the HTTP request).
Also, the browser needs to check that the $\textURL$ value in the
manifest corresponds to the web application's URL (excluding the query part of the URL).

Providing a signed manifest indicates the website (i.e. the URL) opted into the protocol.
From now on, the extension will expect an accountability
manifest until the users explicitly choses to opt
out.

Apart from the manifest generation, 
the signing operation and uploading the signature to the ledger
can also be automated 
thanks to existing tool support for \ac{SRI} and \ac{SXG}.
We stress that the signatures need only be computed if the \ac{JS} code
changes. Techniques like microtargeting are disincentivised by
accountability (see \cref{sec:threat-model}), hence the performance of the
signature generation is of secondary concern.

%% file: protocol.tex
\section{~~~~Protocol}\label{sec:protocol}

\input{protocol-figure}


In this section, we present the Accountable JS protocol.
The end-to-end goal is to hold the developer accountable for the
active content the client receives.
Clients can compare this code with the manifest, hence, for honest
clients, we can reformulate this task as follows:
\begin{itemize}[leftmargin=1.2em]
    \item Clients should only run active content that follows
        the manifest. This is a setup assumption.
    \item Any manifest the client accepts needs to originate from the
        developer, even if the developer or server is dishonest. This
        follows from the non-repudiation of origin property of the
        signature scheme. A signed manifest was either signed by the
        developer, or the developer leaked their key.
    \item Whenever two clients accept a manifest with the same version
        number, that manifest must be the same, or they can provide
        non-repudiable proof that this was not the case. This is
        achieved by including a transparency log that gathers all
        manifest files with valid signatures.
    \item Whenever a client accepts a manifest with some version
        number, this version was the latest version in some
        client-defined time frame. This is achieved by a timestamping
        mechanism like \ac{OCSP}-Stapling~\cite{ocsps2011}.
    \item A client can provide non-repudiable proof that they received
        a manifest from the web server. This is achieved by signing
        a client-provided nonce.
\end{itemize}

The developer of the website generates a manifest file for the 
    web page that is identified with a URL, signs the manifest and 
    publishes it in one or more public transparency logs 
    \begin{nnew}(see Fig.~\ref{fig:protocol-flow} before $t$)\end{nnew}.
The signature proves to the client that the developer takes
responsibility of the manifest.

The \CodeStapling protocol ensures that, whenever the client accepted
a manifest, the developer can be held accountable for publishing it.
Nevertheless, the developer cannot be held accountable for delivering
it to a client, as there is no proof for that.
We thus define the \CodeDelivery protocol 
for non-repudiable code delivery 
\begin{nnew}(in Fig.~\ref{fig:protocol-flow} after $t$)\end{nnew}.
With the HTTP GET request, the client $C$ sends a nonce $\nonce$
signed with its signing key $\sk_{C}$.
The web server $W$ responds with a signature on 
     the HTTP response \HTML,
     the client nonce $\nonce$,
     and signed log timestamp $\siglog$.
The client validates the log's signature and the developer's signature
within. Should one of these checks fail, the client aborts and
displays an error message.
Then, the client compares the active content in $\HTML$ with the
manifest; if they are consistent, the browser decides the web page
adheres to the protocol.
%

%% file: protocol-figure.tex
\begin{figure*}[h!]
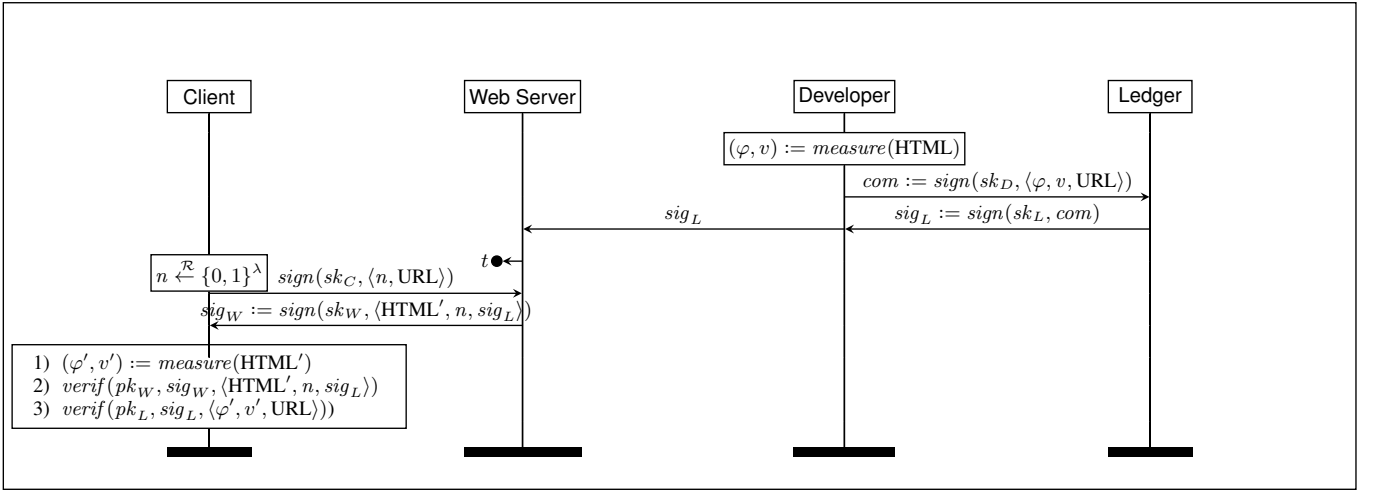
 
\vspace{-0.8cm}
\centering
\resizebox{\textwidth}{!}{
\setmsckeyword{} 
\begin{msc}[small values,
    /msc/level height=5mm,
    /msc/label distance=0.4ex, 
    /msc/first level height=0.3cm,
    /msc/last level height=0.4cm,
    /msc/environment distance=32mm
    ]{}
\setlength{\instwidth}{1.3\mscunit} 
\setlength{\instdist}{3.3\mscunit} 
\declinst{c}{}{Client}
\declinst{w}{}{Web Server}{}
\declinst{d}{}{Developer}{}
\declinst{l}{}{Ledger}{}

\action*{$(\measurementInSignature,v)  \defeq \Measure(\HTML)$ }{d}
\nextlevel[2]
\mess{$\com \defeq \Sign(\sk_D, \langle \measurementInSignature, \version, \textURL \rangle) $}{d}{l}
\nextlevel
\mess{$\siglog \defeq \Sign(\sk_L, \com) $}{l}{d}
\mess{$\siglog $ }{d}{w}
\nextlevel
\lost[side=left]{}{$t \,$}{w}
\action*[fill = white, yshift=1mm]{$\nonce \rdraw \bin^\secpar$}{c}
\nextlevel
\mess{$\Sign(\sk_C, \atup{\nonce, \textURL})$}{c}{w}
\nextlevel
\mess{$\sigtuple \defeq \Sign(\sk_W, \atup{\HTML', \nonce, \siglog})$}{w}{c}
\nextlevel
\action*[yshift=2mm]{
    \parbox{6cm}{ 
    \begin{enumerate}
        \item $(\measurementInSignature',v')  \defeq \Measure(\HTML')$
        \item  $\VerifySign(\pk_W, \sigtuple, \atup{\HTML', \nonce, \siglog})$
        \item $\VerifySign(\pk_L, \siglog, \atup{\measurementInSignature',v',\textURL})$)
    \end{enumerate}
}
}{c}
\nextlevel[2]

\end{msc}
} 
\caption[caption]{
    \begin{nnew}
    Protocol flow: $\CodeStapling$ (before t) and $\CodeDelivery$ (after t).
    \end{nnew}
 }\label{fig:protocol-flow}.

\end{figure*}








%% file: verification-intro.tex
\section{~~Protocol Verification}\label{sec:verification-intro}

We analysed Accountable JS with Tamarin~\cite{meier2013tamarin}, considering the protocol's security
w.r.t.\ a Dolev-Yao adversary that can manipulate messages in the network and 
corrupt other processes to impersonate them. Using Tamarin's built-in 
stateful applied-$\pi$ calculus~\cite{kremer2014automated}, we could model a global state such as represented by the transparency
log.
%

The protocol comprises five processes running in parallel:
\[
    !P_\mathit{Developer} \mid !P_\mathit{Webserver} \mid !P_\mathit{Client} \mid !P_\mathit{Log} \mid P_\mathit{Pub} 
\]
The first three processes model the role of the developer, web server and
client, outputting and accepting messages as specified in
\begin{nnew}
Figure~\ref{fig:protocol-flow}.
\end{nnew}
%
%
%
The developer, web server and the client are under replication to account for an unbounded number
of parties acting in each role. 
\begin{nnew}
Any party except the log and the public process can become dishonest.
This is modelled by
giving control to the adversary, but only after emitting
\fixed{\reviewnote{R\ref{r1}.\ref{r1:extend-protocol-verification}}}
a $\mathit{Corrupted}$ event, which can be used to distinguish the
party's corruption status in the security property.
A corrupted party remains dishonest for the rest of the protocol execution.
\end{nnew}
%


\begin{nnew}
The process $P_\mathit{Log}$ models an idealised append-only log
using insert and lookup operations to a global store~\cite{kremer2014automated}.
Moreover, the built-in lock and unlock commands are used to ensure
atomicity of the operations. 
Finally, the process $P_\mathit{Pub}$ make the public's
ability to validate a client's claim explicit.
\fixed{\reviewnote{R\ref{r1}.\ref{r1:extend-protocol-verification}}}
Upon obtaining a claim (from the client), this process 
(1) reads from the log the information that concerns the URL mentioned
in the claim,
(2) verifiers the signatures in the claim
and
(3) matches the signed values with those in the log.
\end{nnew} 

Using Tamarin, we prove the following properties which are detailed in Appendix~\ref{sec:protocol-verification}. 
\begin{itemize}[leftmargin=1.2em]
	\item \textbf{Authentication of origin}: The client executes active content only if 
	the corresponding manifest was generated by the honest developer unless the adversary corrupts the developer. 

	\item \textbf{Transparency}: If the client executes code then its manifest is present in a transparency log in a sufficiently recent entry.
	
	\item \textbf{Accountability}: When the public accepts a claim, then even if the client was corrupted, 
            the code must exist in the logs and the server must have sent that data (either honestly or dishonestly via the adversary).

    \item \textbf{End-to-end guarantee} : Only by corrupting the developer it is possible to distribute malicious code.

\end{itemize}

%% file: code-verify.tex
\subsection{Code Verify Protocol}\label{sec:code-verify}

Meta's Code Verify~\cite{codeVerify} was published in March 2022 and
made available as an extension. As of now, it is deployed only by
WhatsApp Web. 
Intuitively, WhatsApp Web (the developer) submits a hash of their JavaScript along with a version number to Cloudflare, which 
Cloudflare then publishes to the end user.
The end user's browser extension computes a hash on the JavaScript delivered from WhatsApp Web and 
compares it  against the hash published by the Cloudflare. 
Given that the manifest is hashed instead of signed, Cloudflare is
trusted for authenticity and thus constitutes a trusted third party, replacing the log.
\begin{nnew}
Moreover, users' IP addresses are sent to Cloudflare instead of to WhatsApp Web.
\end{nnew}

We likewise modelled Code Verify in Tamarin, considering the following
five processes:
\[
    !P_\mathit{Developer} \mid !P_\mathit{Webserver} \mid !P_\mathit{Client} \mid !P_\mathit{Cloudflare} \mid P_\mathit{Pub} 
\]
Again, we assume the developer is separate from the web server. 
\begin{nnew}
The protocol does not have a public log and does not include independent auditors.
\fixed{\reviewnote{R\ref{r1}.\ref{r1:discuss-code-verify}}}
Instead, Cloudflare records the hashes for each version.
To our knowledge, Cloudflare
does not provide information about 
the history of submitted versions
or
which is most recent.
As the public cannot inspect how often versions have changed,
it relies on Cloudflare to implement countermeasures against
\fixed{\reviewnote{R\ref{r4}.\ref{r4:critics-meta-proposal}}}
microtargeting. Publicly available information~\cite{codeVerify} did
not give information about such measures in Meta's deployment. 
\end{nnew}

Under these considerations, we analysed the same properties, except
\begin{nnew}
for transparency, which, due to the lack of
\fixed{\reviewnote{R\ref{r1}.\ref{r1:discuss-code-verify}}}
a public log, could not apply. We highlight the differences to our original properties below.
\end{nnew}

\begin{itemize}[leftmargin=1.2em]
	\item \textbf{Authentication of origin}: The client executes active content only if 
	the corresponding manifest was generated by the honest developer unless the developer \underline{or Cloudflare} is corrupted. 
	
        \item \textbf{\underline{Non-}Accountability:} The data provided to the
            client is not sufficient to prove they received certain
            content from the web server, even if web server and
            Cloudflare are honest.

        \item \textbf{End-to-end guarantee}: Only by corrupting the
            developer \underline{or Cloudflare} it is possible to distribute
            malicious code. In a separate lemma we show that,
            the developer by itself can indeed distribute malicious
            content.

\end{itemize}

The latter property indicates that Cloudflare's role as trusted party
is not fully exploited yet.
\begin{nnew}
At least as far as we know~\cite{codeVerify},
Cloudflare neither promises to ensure the code is harmless,
nor does it guarantee to collect information to provide
transparency or accountability.
\fixed{\reviewnote{R\ref{r4}.\ref{r4:critics-meta-proposal}}}
Nevertheless, 
the current message flow can
be extended to provide such guarantees by having Cloudflare acts
as a transparency log.
\end{nnew}
Accountability can likewise be achieved by simply deploying signatures
instead of a hashing scheme. 
%

%% file: transparency-log-intro.tex
\section{~~~Logging Mechanism}\label{sec:transparency-logs}

We would like clients to verify they received the latest and same version of the code as any other user.
To this end, we assume a public append-only log to provide a public record of the
software published and prevent equivocation attacks.
The log does not determine which \ac{JS} is considered
malicious, but it provides proof of receipt and origin, and allows
identifying malicious versions.

Such a public log is realistic to deploy:
\ac{CT} Logs~\cite{rfc6962} are used in the modern internet infrastructure.
These logs store certificates, which are signed by \acp{CA}.
In contrast, our logs need to store manifests signed by the developers.
It is thus impossible to reuse the existing \ac{CT} infrastructure,
but we can closely follow the structure and properties of \ac{CT}.

Websites that offer security-conscious services have an incentive to retain their reputation.
Similar to how \ac{CT} logs operate, our log can be run by a party that wants to support such webpages.
Third party monitors can keep the monitor honest and we allow third parties to submit signed manifests
they observe in the wild.

When implemented naively, a logging mechanism can bring significant privacy implications:
To confirm that other clients
receive the same manifest, the client would need to consult the log on
each request and reveal the URL to the log.
We can mitigate these privacy issues by
allowing the web server, which learns each request anyway, to
include a signed and timestamped inclusion statement from the log instead.
This is similar to the \ac{OCSP}-Stapling for 
certificate revocation status requests
~\cite{rfc6961}.
While it mitigates the privacy issues of consulting the log,
it instead requires the user to trust the specific log selected by the web server.
We outline other approaches to solve the trade-off between trust and privacy in \cref{sec:related}.



Overall, our transparency log needs to provide
interfaces to at a minimum:
\begin{itemize}[leftmargin=1.2em]
    \item store the signed manifest file (including its version number)
        bound to a URL,
    \item query the latest signed manifest file for a URL in the logs, 
    \item form a signed response for a query 
        that can be pre-fetched by the web server to staple it 
        to each request from the clients.
\end{itemize}
A possible implementation of this functionality could be based on Verifiable Log-Based Maps~\cite{verifiableDataStructures}. An implementation of this structure for Trillian~\cite{trillian}, the software running Google's \ac{CT} server, is currently in progress~\cite{batchmap}, with the goal of supporting transparency in certificate revocation\fcite{revocationTransparency}.
%

%% file: transparency-logs.tex

Availability, scalability and the size of the transparency logs are other implications.
Be it submitting a new manifest to the log or collecting the latest version of manifest for a URL, 
low latency to access the network of transparency logs
%
can be achieved by 
eliminating the single point of failure by adding multiple logs that will provide load balancing.
%
The mechanism proposed for query privacy will also decrease the number of requests to the logs since the web server will provide the stapled result in most cases.

Websites that frequently update their active contents can create significant burden on the log size. 
We calculate approximately how many times each log can be updated for a limited time and space.
We assume a non-leaf node overhead is approximately 100 bytes and for the leaf nodes it is 700 bytes(signature 600 bytes + 100 bytes).
If a log provider has 100 TB of space for 5 years, it can contain 137 billion signatures in total.
%
To make sense of this number, take the following example. We start
with a log of 10M URLs with eight updates per month on average.
The number of URLs also increases exponentially at a rate of 1\% with
each update (i.e. also eight times per month). 
\footnote{e.g. after the first update,
10M updates along with 100k new URLs are appended to the existing 10M,
resulting in a total of 20.1M.} This number would be well below 137
billion signatures.

%% file: eval.tex
\section{~~~~Evaluation}\label{sec:eval}

\input{eval2}

We implemented Accountable JS in a Chrome extension
\cite[folder \texttt{accjs-extension}]{supplementary-files}
for demonstration and prototyping.
Ideally, the measurement procedure should be part of the browser's
rendering engine, since it can access the response body and
observe mutations to elements first-hand.
Our measurements here can thus be (promising) upper bounds.
We elaborate on the technical limitation imposed by the extension
SDK in Section~\ref{sec:limits_prototype}.
%
%

%
We come back to the use cases from \cref{sec:case-studies} and
measure how the extension affects the following
metrics:
\begin{enumerate*}
    \item number of additional requests,
    \item size of additional traffic,
    \item time until the browser paints 
         the first pixel
       / the largest visible image or text block\footnote{More precisely: the `largest contentful paint'.}
       / until the web page is fully responsive.
    \item total blocking time, i.e. time during which web page cannot process user input.
\end{enumerate*}
We consider differences below 100 ms to be imperceptible to the users,
differences of 100-300 ms barely noticeable 
and differences above 300 ms noticeable.%
\footnote{
We derive these performance categories from the RAIL
model~\cite{webdev-rail}. 
%
According to RAIL, users feel the result is immediate if $<100$ ms and
feel they are freely navigating between 100-1000 ms (see also~\cite{response-times}).
However, we found this gap is too wide to ignore,
and split the category at
300 ms 
for an unusually common delay in web apps due to the
`double tap to zoom' feature on iPhone
Safari~\cite{what-is-300ms-delay}.%
\processifversion{marked}{
(R\ref{r2}.\ref{r2:performance-ranges})
}
}

\paragraph{Evaluation environment}
Measurement took place
on 
a MacBook Pro with 2 GHz Intel Quad-Core i5, 16 GB RAM and
macOS Monterey 12.5.1
with
Google Chrome 107.0.5304.121.
%
The results are compiled in Table~\ref{tab:eval-results}. 
We measured the number of additional requests and traffic using
Chrome's developer tools and 
the rendering metrics using
Lighthouse (set
to 
`desktop simulated throttling'). 
Unfortunately, WhatsApp Web is incompatible with Lighthouse,
so we instead 
computed the 
combined duration of all tasks performed by the browser
using Puppeteer Page metrics~\cite{Page-metrics}. 
%
We automated this process using Puppeteer and NodeJS and perform 
$n=200$ trials per website and configuration to minimise the impact of network latency on page loads.

\paragraph{Configurations}
For performance evaluation, we compare the \ac{CSP} built into the browser with the Code Verify and Accountable JS extensions as follows: 
\begin{enumerate}
\item \textbf{Baseline: } disabled \ac{CSP} and extensions.
\item \textbf{\ac{CSP}: } \ac{CSP} active, no extension. 
\item \textbf{Accountable JS: } \ac{CSP} inactive, only Accountable JS extension active.
\item \textbf{Code Verify: } \ac{CSP} inactive, only Code Verify extension active. This configuration only applies to 
    WhatsApp Web, as Code Verify currently only supports Meta websites.
\end{enumerate}

\paragraph{Experiments}
We consider the examples from \cref{sec:case-studies}:
Hello World, WhatsApp Web,
Trusted Third-Party,
Delegate Trust to Third Parties (Nimiq A), Untrusted Third Party (Google AdSense and Nimiq B).
For the compartmentalisation experiment on Nimiq's Wallet,
we use a different baseline that we will discuss below.
For the \ac{CSP} measurement,
we defined \ac{CSP} headers for each website
that listed all active content in the
Accountable JS manifest files. 
We collected all valid sources of external scripts and 
hashes for the external and inline scripts in \ac{CSP}'s \texttt{script-src} directive,
hashes for event handlers in \texttt{script-src-attr} 
and
sources for iframes in \texttt{child-src}.
For the Accountable JS experiment, we first navigate to the target website and wait for ten
seconds for the page to load.
Thereafter, using the $\generateManifest$ in the extension, we download the manifest file and self-sign it 
using the \gensxg command line tool~\cite{Signed-exchange-code}. 
For Nimiq A+B and AdSense, 
we changed the $\mantrust$ attribute for the external element(s) to
$\delegate$ before signing.
We publish this signed manifest via a local web server and configure the web server to provide a response header 
pointing to a $\textURL$.
We also ensure the website provides \ac{SRI} tags for $\external$ scripts.
Evaluation procedures of each case study are elaborated in
\theappendixorfull{sec:evaldetail}.


%

%
%
%

\paragraph{Results}
The \ac{CSP} configurations show an imperceptible overhead in all case studies.
This is hardly surprising, as \ac{CSP} is built into the browser built-in and can
validate
resources during rendering.
A detailed \ac{CSP} defined for Nimiq A (Nimiq including its own \ac{CSP})
increases the reaction time by about 65 ms. 
The Accountable JS configurations likewise have an imperceptible
overhead
\begin{nnew}
in all case studies.
\end{nnew}
%
Moreover,  the traffic requirements are modest and incur only modest
blocking time. 
For Nimiq A, the traffic requirements are about 9.9 kB for the additional
signature. In terms of performance, \ac{CSP} and Accountable JS' overhead are comparable.
The time to interactive value unexpectedly 
\begin{nnew}
decreases 
more with Accountable JS than \ac{CSP}.
\end{nnew}
However, the difference is minimal and could possibly be explained by
a) network latency,
(b) side effects of the browser's just-in-time compilation or
scheduling
or
(c) a side effect of the former two on how Lighthouse evaluates the
reactive metric.
Nimiq is a complex web application heavily dependent on external data,
in particular the remote blockchain it connects to.

\paragraph{Discussion}
The Accountable JS configurations have an imperceptible overhead 
\begin{nnew}
which is slightly higher than the \ac{CSP} configurations.
Recall that the \ac{CSP} is built in the rendering engine 
whereas Accountable JS runs as a browser extension. 
Accountable JS has
\end{nnew}
to perform 
signature validation,
meta data collection 
and
a final compliance check.
The prototype achieves 
\begin{nnew}
good performance overheads
\end{nnew}
by measuring all elements 
simultaneously and combining their results.
The browser extension panel displays the results
instantaneously, while the evaluation is in progress, although the
evaluation is usually too quick for the user to notice.
Moreover,  the traffic requirements are modest and incur little
blocking time. 

%
For AdSense + Nimiq B, the network overhead is slightly higher than Nimiq A.
This is due to the larger size of the
manifest, which now also includes AdSense. 
We again observe an imperceptible impact on performance with Accountable JS.
%

The difference between 
\begin{nnew}
Code Verify (220ms) and Accountable JS (244ms) on WhatsApp Web is
small. 
\end{nnew}
This is remarkable, because
Code Verify only applies \ac{SRI} checks on external scripts but not
event handlers or iframes.
In contrast to Accountable JS, the order of active elements is ignored, 
attributes are not checked (e.g. \texttt{load='async'} for scripts) and
a short hash value is downloaded from Cloudflare, rather than
a signature.

\paragraph{Compartmentalisation}
For compartmentalisation, we evaluate the impact of the additional
signing key. 
We signed Nimiq Keyguard, which is embedded in Nimiq Wallet, with a different signing key
and set the Keyguard's $\mantrust$ attribute to $\delegate$ in the Wallet's manifest.
The baseline therefore also has the  Accountable JS extension activated, but
uses the same signing key on all Nimiq components.
The Wallet's manifest includes the Hub's manifest inside and the Hub's manifest declares 
the Keyguard with $\mantrust=\delegate$ in its manifest section. 
Thus a separate
manifest is required for the KeyGuard.
Also, this time a separate signing key is used for the KeyGuard manifest.
For the baseline performance, we inline the KeyGuard’s manifest as
an entry for its iframe in the Wallet's manifest, thus having one
manifest and one signing key, and activate the extension.

In the compartmentalisation evaluation, we observe that there are two more round trips and slightly higher
traffic overhead (about the overhead of Accountable JS, w.r.t. the
overall page traffic of 4.6 MB). 
This is due to downloading the extra \ac{SXG} certificate and manifest for Keyguard.
The effect on the rendering metrics is small; 
\begin{nnew}
the barely noticeable increase for time-to-reactive value can again be explained 
with network latency and side effects described above.
\end{nnew}
This is because the delegated manifest can
be validated in parallel to rendering, while it is inlined in the
baseline scenario and thus validated in sequence.


Due to stapling, the overhead for clients to verify that they received
the latest version of the code (and thus the same as any other user), 
is negligible. 
The web server staples a query result, i.e. the log's signature on
the signed manifest, to each request.
The signatures use 2048-bit RSA keys and are
256 Byte long.

%% file: eval2.tex
\begin{table*}[hbt!]
\caption{Evaluation results on case studies\label{tab:eval-results}:
    The second and third columns show the number and total size of additional requests made by
    the extension, i.e. the number of signed manifest and certificate. 
    Each subsequent block provides Lighthouse performance
    metrics for rendering time and the total time that the browser
    spends unresponsive.
    For each metric, we compare
    the baseline (no \ac{CSP}, no Accountable JS)
    with the overhead incurred by enabling \ac{CSP}
    and enabling the Accountable JS extension (leaving \ac{CSP}
    disabled). 
    For compartmentalisation, the baseline is with the extension
    activated but the same signing key for all Nimiq components.
    %
    All the time values are averages over $n=200$ runs and
    given in milliseconds.
    The additional traffic(kB) value is affected by the size of 
    the signature and \ac{SXG} certificate. Signatures
    are generated on uncompressed manifest JSON files.
}
\centering
\begin{tabular}{lccrr@{}rrrr@{}rrrrr@{}rrrrr@{}rrrrr} 
    \toprule
    & 
    \multicolumn{2}{c}{additional network \ldots} &
    \multicolumn{17}{c}{time to \ldots baseline $+$ \ac{CSP} overhead $+$ Accountable JS overhead} 
    \\
        \cmidrule(r){2-3}
        \cmidrule(r){6-23}
 case study &
 requests  &
 traffic (kB) &
 \multicolumn{5}{c}{first pixel} &
 \multicolumn{5}{c}{largest element} &
 \multicolumn{5}{c}{reactive} & 
 \multicolumn{5}{c}{blocking time} 
 \\
 \midrule

Hello World & 
2 & 2.06 & & & 196 & +1 & +20 & & & 197 & +0 & +23 & & & 196 & +1 & +24 & & & 0 & +0 & +0 
\\
Trusted Third-Party &
2 & 2.46 & & & 462 & +0 & +21 & & & 462 & +0 & +21 & & & 462 & +0 & +21 & & & 0 & +0 & +0
\\
Delegate Trust (Nimiq A) &
3 & 9.93 & & & 262 & +3 & -10 & & & 262 & +3 & -10 & & & 5591 & -29 & -144 & & & 172 & +4 & +87 
\\
AdSense + Nimiq B &
3 & 15.62 & & & 747 & +2 & +91 & & & 901 & +5 & +68 & & & 6034 & +1 & -82 & & & 159 & +3 & +77
\\
\midrule
Compartmentalisation &
2 + 2 & 8.66 +1.10 & & & 2200 & & -17 & & &  4675 & & +20 & & & 5321 & & +115 & & & 212 & & +7
\end{tabular}
\vspace*{-2mm}
\end{table*}

%% file: limitations.tex
\section{~~~~Limitations of Prototype}\label{sec:limits_prototype}

The browser extension is a prototype to evaluate performance and
applicability of the approach.
The advantage of an extension (as opposed to
modifying the browser’s source or writing a developer plugin) is that
users can easily experiment with its code.
On the other hand, extensions cannot
interrupt the browser's rendering engine. Thus we
inject a content script\fcite{CONTENT-SCR} that can apply
the client-side operations of the protocol to the browser window.
The content script runs in the same context as the web page; hence it can observe changes to the \ac{DOM} via the Mutation Observer.
Since the extension cannot access to the browser's rendering engine,
some active elements can be added within
a small time frame before
the Mutation Observer is registered.
This race condition is a limitation of using the extension and fixable
by closer integration into the browser.

Another limitation is that other browser extensions may interfere with
the measurement by injecting active content to the web page. Since
extensions cannot distinguish website code from the code that other
extensions injected to the web page, this can break the measurement.
This is the correct behaviour, as the website developer cannot attest
to every possible modification of the active content by other extensions, however, there
are various client-side solutions: (a) closer integration into the
browser could distinguish active content injected by websites, (b) the
extension could provide an API for third party extensions to register
modifications or (c) an allowlisting for the most common extensions
that gives a warning to the user.

%% file: related.tex
\section[Related Work]{~~~Related Work}\label{sec:related}

We first discuss how Accountable JS relates to other (proposed) web standards with
seemingly similar goals, before discussing related academic proposals.

\ac{CSP} was introduced to counter \ac{XSS} attacks. They specify runtime
restrictions for the browser, typically the set of allowed sources for
scripts, iframes, stylesheets, etc., including eventual requirements for
sandboxing.
Like accountability manifests, \acp{CSP} can specify which sources are
allowed and, combined with \ac{SRI}, fix their content. This is
comparable to a manifest file that includes types with \mantrust set
to either \assert (if \ac{SRI} is employed) or \blindtrust
(otherwise). 
By contrast, \acp{CSP} do neither cover the order nor possibly nested
active contents (e.g. iframe within iframe). Mixed ordering of active content may create malicious activity, 
a site loading script A before script B may mean something different from loading B before A. 
A site that only uses \ac{CSP} cannot catch that behaviour, whereas in Accountable JS, we take the order into account.
Most importantly, in \ac{CSP}, there is no means of delegating trust and no
distinction between web server and developer.
Steffens et al.~\cite{steffensWhoHostingBlock2021} 
show that outsourced content is one of \ac{CSP}'s major deployment
obstacles.
Instability in third party inclusions (e.g. adbidding code that
delivers code from different resources) 
forces first parties to continuously update the \ac{CSP}.
Techniques like in \cref{sec:adsense-use-case} allow developers
to delegate trust to the third party.
\begin{full}%
It is conceivable to incorporate features like trust delegation into
\ac{CSP}, along with a mechanism for signing \ac{CSP} headers. By contrast,
capturing the order or the nesting between active elements and providing
non-repudiation appears to clash with the design of \ac{CSP}.
\end{full}%
Moreover, \ac{CSP} tries to mitigate \ac{XSS} attacks throughout the web, whereas 
Accountable JS targets websites willing to allow for an audit. 
The ability to identify the code that is run is a key requirement for that. 
Overall, the goals of \ac{CSP} and Accountable JS are orthogonal and
can be combined. It is possible to generate a \ac{CSP}
from a manifest file.

\begin{figure}[b!]
\centering
\includegraphics[width=\linewidth]{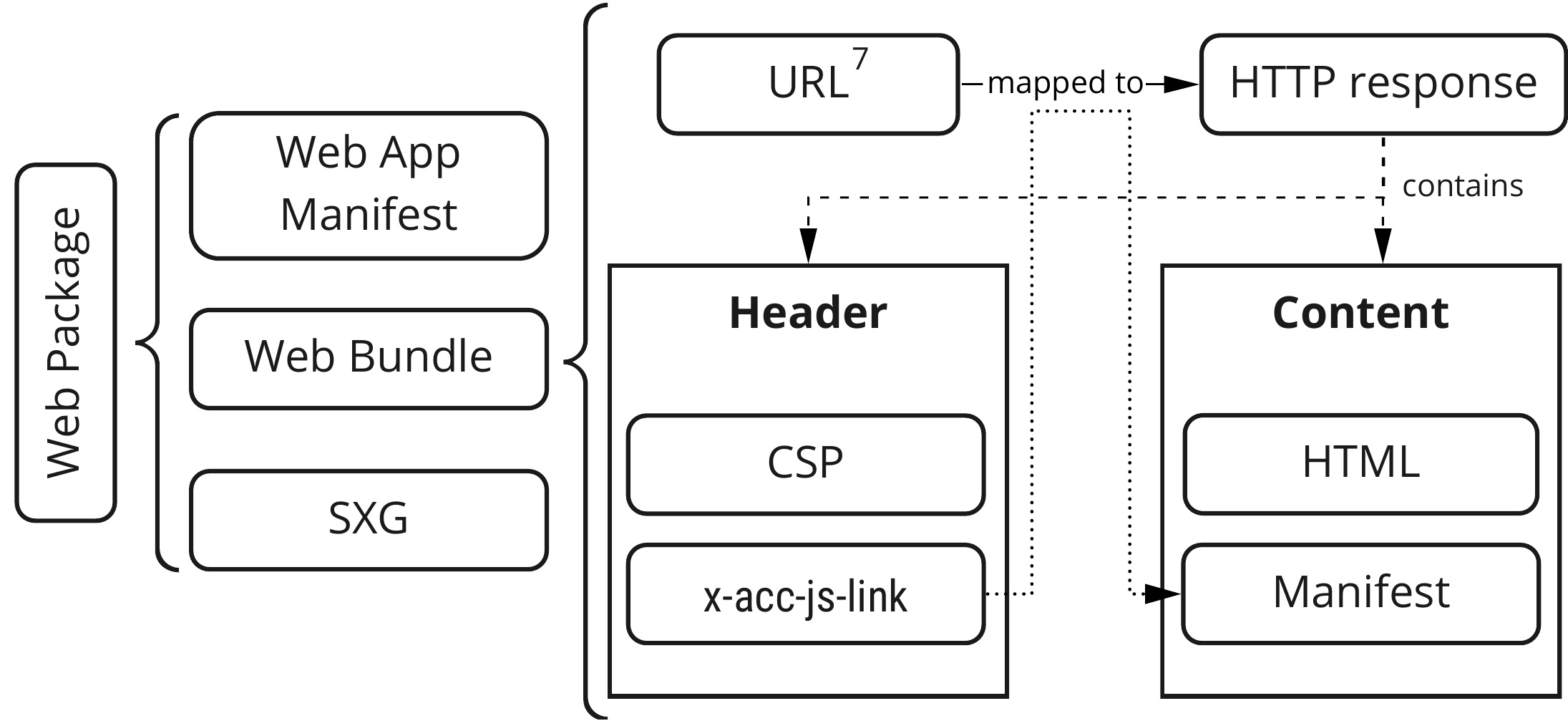}
\caption{Accountable JS in the context of other web technologies.}
\label{fig:relationship-between-web-standard-and-accountable-js}
\vspace*{-5mm}
\end{figure} 

The Web Package proposal (currently in draft status~\cite{yasskinUseCasesRequirements}, 
see \cref{fig:relationship-between-web-standard-and-accountable-js})
aims at packaging web
applications for offline use.
Web packages provide
a declaration of the web application's metadata via \emph{Web App Manifests}~\cite{marcoscaceresWebAppManifest}, 
a serialisation of its content via \emph{Web Bundles}~\cite{yasskinWebBundles},
and authenticity via \ac{SXG}~\cite{SXG-Webpackage}.
We likewise employ \ac{SXG} to provide authenticity of origin via
signatures.\begin{nnew}%
    \ac{SXG}, like Accountable JS, decouples web developer from website hoster.\fixed{\reviewnote{R\ref{r3}.\ref{r3:sxg-more}}}
\end{nnew}
Web App Manifests, despite their name, are only superficially related.
They contain startup parameters like language settings, entry points
and application icon, e.g. for `installable web application'
displayed in a smartphone's launcher.  
Web Bundles are a serialisation format for pairs of URLs\footnote{More
precisely, HTTP~\emph{representations}~\cite{fieldingHTTPSemantics}.}
and HTTP responses.
They represent a web application as a whole, but a signature on a web
bundle would change with every modification of a web pages' markup.
Web Packages are thus not competing with Accountable JS. Instead, both
standards are compatible. A web bundle can contain 
\accjsheader in the header of its entry point's HTTP response,
triggering the browser to validate the manifest. The manifest is
specified via a URL that also included in the web bundle. This URL
maps to an HTTP response that contains the manifest in its content
part.

Signature-based \ac{SRI}~\cite{westMikewestSignaturebasedsri2020} 
proposes easier maintainable \ac{SRI} tags to protect against script injections, by including
signature \emph{keys} instead of hashes. These enable validating the provider
of the third party script, instead of their content, like 
the trust relationships expressed with $\mantrust=\delegate$.
The tags are part of the HTML code, instead of the manifest file.
Signing the HTML files is impractical, as they are frequently changing.

Service Workers~\cite{russellServiceWorkers} are Network proxies
programmable via JavaScript, often used to perform URL response
caching, separate from the browser cache. Theoretically, a compliance
check like our measurement could be implemented in a service
worker, but (a) the service worker would need to be delivered
correctly and (b) service workers lack access to the \ac{DOM} and thus
information about how active elements used.

We will now discuss related academic work.
Accountability in the web requires non-repudiable proof. For static
assets, this can, in principle, be provided by digital signatures
(e.g. via \ac{SXG} and web bundles, see above), but recreating the
signature for each exchange is costly. 
We solve this via a simple challenge-response mechanism.
Ritzdorf et al. \cite{ritzdorfTLSNNonrepudiationTLS2018} provide a 
full-fledged
solution, giving non-repudiation for the entire communication,
optionally hiding sensitive data.
The statement we prove is that the client has \emph{obtained} certain
active content, not that they \emph{execute} it. Ensuring a remote
partner runs certain software is the goal of remote code attestation
(e.g. \cite{guRemoteAttestationProgram2008}).
Outside embedded systems, this is typically based on a trusted
execution environment (e.g. TPM, SGX). While the browser (and for
that matter, our extension) could provide a trusted execution,
establishing trust in the correctness of the browser is the crux.

Our work relies on a transparency log. As mentioned before,
Trillian's~\cite{trillian} verifiable log-based maps would fit the
bill, but there are many ways to implement such a store. The most
interesting aspect is privacy. We propose an approach based on
stapling, an established method for revocation
management~\cite{ocsps2011}, but other techniques promise
privacy, too. CONIKS~\cite{coniks190974} provides a log mapping
user identities to keys, keeping the list of all user identities in
the system private. This is not useful in our case, as the URLs (the
domain of our mapping) are not secret, but which URL a user accesses.
Multiparty protocols for Private Information
Retrieval~\cite{fortnowPrivateInformationRetrieval}, Private Set
Intersection (e.g. \cite{pinkasSpOTLightLightweightPrivate2019}) or
ORAM~\cite{goldreichSoftwareProtectionSimulation1996} lack efficient
database updates, mechanisms to efficiently update precomputation
steps, or only preserve k-anonymity for URLs.
K-anonymity is  often not enough if we consider that an attacker, e.g.
a censor, tries to punish access to a few critical URLs, each of which
may end up in a bucket with uncritical, but also not frequently
visited URLs. 
%
Finally, Accountable JS may be an enabler for formal verification of
web applications, as users are potentially able to link the code they
receive to code to published verification results. Various static and
dynamic analyses target JavaScript already~\cite{parkKJSCompleteFormal2015,santosSymbolicExecutionJavaScript2018}.

Although we showcased only a single approach to code compartmentalisation (as it is being deployed by our real-world example), other approaches are also compatible with Accountable JS.
Language-based isolation methods like BrowserShield~\cite{reis-2007-browsershield}) rewrite JavaScript into a safer version preventing or mediating access to 
critical operations like \texttt{createElement} or \texttt{eval}.
If the code is rewritten on the client (typically using
JavaScript), the developer declares the wrapper that
fetches the code and deals with the code rewriting in the manifest
file.
If the code is rewritten on the server, the developer declares the
transformed JavaScript code that will be delivered to the user.
Frame based isolation methods (e.g. AdJail~\cite{267210}) that isolate the third party code inside iframe are also compatible with our proposal, see the use case for untrusted third party code in Section~\ref{sec:adsense-use-case}.

%% file: discussion.tex
\section{~~~~Discussion}\label{sec:discuss}

We provide a solution that allows users to detect if 
they are microtargeted by developers and to prove this to the public if it is the case.
Sending different codes to classes of users might not 
be outlawed in many countries, but sending malicious code is.
Our solution neither provides a code audit tool nor does it propose
a framework, legal or otherwise, for the punishment of malicious code
distribution. 
It provides, however, verifiable data that authorities can use to
evaluate which code was published and whether that code was delivered
to a user. 
Moreover, the protocol provides users with a claim that includes the 
delivered code and the identity of the developer.

The transparency logs can be used as a point of reference for the
public code for auditing and evaluating. Honest developers aim to make
their code easy to audit; dishonest developers thus risk loss of
reputation if they 
microtarget users (as frequent updates are visible on
the transparency logs),
silently opt out of the system (as this will be caught by users that received a previous opt-in),
or provide obfuscated code (due to the log).

Honest developer will benefit from a good reputation and their ability to
provide proofs for any efforts they make toward independent audit or
formal verification.
Clients, who often debate a website's reputation in a public forum
(e.g. the case of ProtonMail or Lavabit) obtain data to substantiate positive and negative claims.

\begin{nnew}
We stress that accountable code delivery is necessary to connect
auditing results to the code users actually run, but does not by
itself guarantee the safety of this code.
Realistically, it will take some time until software analyses are
mature enough to handle this at scale. Assuming, however, that such
analyses may not necessarily run at each browser independently,
authentic code delivery appears to be a necessary first step.
\end{nnew}

\begin{nnew}
Moreover, Accountable JS only authenticates the active content,
thereby exposing the active content to data-only attacks, e.g.
modified button labels or form URLs.
A signature on the content of a web application could be achieved by
building on Web packages/Web bundles (which we discussed
in \cref{sec:related}), however, this approach would be too static and
inflexible for the requirements of the current web ecosystem.
\fixed{\reviewnote{R\ref{r4}.\ref{r4:non-active-content}}}
Thanks to accountability, the developer would take responsibility for
the active content that they published, in this case, for code that is vulnerable
to data-only attacks. Realistically, there would not be consequences,
because they can plausibly point to the dire state of verification of
JavaScript---which is at least partially because users could thus far
not be sure to receive the verified code anyway.
Accountable JS choice to validate the active content only is
a compromise and possible starting point for future work, as we
discuss in the next section.
\end{nnew}

%% file: conclusion.tex
\section{~~~~Conclusion}\label{sec:conclusion}

With Accountable JS, we provide a basis for the accountable delivery
of web applications, and thus a first step towards re-establishing the
trust between a user and the web application code they run on their
computers. How to achieve security – via audit, code analysis
or formal verification – is a question that we left open intentionally.
Accountable delivery is, nevertheless, a requirement for any
non-instantaneous analysis. 

A key question for verification and audit is how 
to relate 
the
web page's user interface to the active content. 
As some desirable security properties concern user input, we would
like to give guarantees about, e.g. form fields. We can account for
the JavaScript code that address them by ID, but those are invisible
to the user. Future work may investigate how to establish stronger
ties between the manifest and the user interface.

\paragraph{Acknowledgements}

This project was partially funded by 
the ERC Synergy Grant IMPACT (with grant agreement 
number 610150) and a research award for privacy-preserving 
technologies from Meta research, specifically 
for the "Transparency.js, transparency for active content" initiative.

%% file: verification.tex
\section{Verification of Security Properties}\label{sec:protocol-verification}

By default,
Tamarin assumes that the adversary controls the network.
Our model allows the adversary to impersonate
the untrusted parties in the protocol and thereby 
access their secrets. This is logged with a $\Corrupted(p)$ event in the trace
with $p$ an identifier for the corrupted principal.

We model the principals in the following structure using applied-$\pi$ calculus.
\begin{lstlisting}[language=sapic]
in($p); 
(
    ( event Corrupted($p); 
      out(sk($p)) )
    | out(pk(sk($p)))
    | (  
        /* process goes here */
      )
)
\end{lstlisting}
The shortcut $\$p$ denotes that the term $p$ is a public value.
The attacker, by inputting the public value $\$p$ can pick some
identifier for the party. Then, if the attacker corrupts the party,
a corruption event is emitted and the attacker gets access to the secret key.
The public (verification) key is emitted so that other parties can use it to verify 
the signed messages of $p$. We exemplify the process with the example of
$P_\mathit{Developer}$ as follows:
\begin{lstlisting}[language=sapic,escapeinside={(*}{*)}]
in($D); 
(
    [..]
    | ( 
        in(<$manifest, $url, $v>); 
        event DUploads($D, $url, (*$\mathit{\varphi}$*));
        out(<'update', $D, $manifest, $url, $v, (*$\mathit{\varphi}$*)>)
        ...
      )
)
\end{lstlisting}
This code snippet includes the interaction with the network via
\texttt{in} and \texttt{out}, which is represented by the attacker.
The attacker hence inputs the
public values $\mathit{manifest}$, $\mathit{url}$ and version number $v$, then
the developer process computes a signature \begin{nnew} $\varphi$ \end{nnew} 
from these values and sends an update
message to the log including all that information. Events are annotations associated with the parts of the processes that enable to define restrictions and security properties. In this example, before sending the update message to the log,
the developer logs a $\DUploads$ event in the trace, annotating the developer's new code update request to the transparency logs. 

\begin{nnew}
The process $P_\mathit{Log}$ represents the transparency log as a protocol party that can receive and send messages, and in addition
apply insert and lookup operations to an append-only global store.
The applied-$\pi$ calculus provides constructs for modelling the manipulation of a global store.
The code snippet below includes an insert and a lookup operation.
\begin{lstlisting}[language=sapic,escapeinside={(*}{*)}]
    insert <$D, $L, 'version', $url>, $v;
    ...
    lookup <$D, $L, 'manifest', $url> as $manifest 
        in P else Q
\end{lstlisting}
The insert construct associates the value $\$v$ to the key which is a tuple $<\$D, \, \$L, \, `\mathit{version}`, \, \$\mathit{url}>$ and
successive inserts overwrite the old values.
The lookup construct retrieves the value associated with the key $<\$D, \, \$L, \, `\mathit{manifest}`, \, \$\mathit{url}>$ and assigns it to $\$\mathit{manifest}$ variable.
If the lookup was successful, it proceeds with process $P$, otherwise with $Q$.
$\$D$ stands for the developer's identity, whereas $\$L$ stands for the log's identity.
Since there are unbounded number of developers and logs; we associate the values that are stored in the global store with
the URL and 
the identities of the related developer and log for uniqueness. 
\fixed{\reviewnote{R\ref{r1}.\ref{r1:extend-protocol-verification}}}

Our model also includes lock and unlock, which the stateful applied-$\pi$ calculus defines for exclusive access to the global store in the concurrent setting. 
The code snippet below shows an example of lock and unlock operations used in our protocol.
\begin{lstlisting}[language=sapic,escapeinside={(*}{*)}]
    lock $url;
    insert <..., $url>, ...;
    ...
    unlock $url;
\end{lstlisting}
When a $\$\mathit{url}$ is locked, any subsequent attempt to lock the same $\$url$ will be blocked until it is unlocked.
We provide exclusive accesses based on the $\$url$, when the log attempts to insert a new value to the global store. This is an over approximation: if a lock requires exclusive access independent for every write (independent of the URL) our model correctly captures this behavior too.
We do not require locks for other reads, which also increases
generality.
\end{nnew}

Security properties and restrictions are first-order formulas over the annotated events and time points.
Universal quantification (meaning: for all) and existential quantification (meaning: there exists) are used to 
check if the security property formula (lemma) holds for all examples in the domain or there exists at least one example
that satisfies the formula respectively. If the lemma holds for the former case then the Tamarin Prover shows that it is proven, whereas 
for the latter case a satisfying example is presented to the user. The time points enable to account for event order in the trace, 
where e.g. $E@i$ means that event $E$ was emitted at index $i$ in the trace.
We prove that the following security properties hold in our protocol:

\begin{theorem}[Authentication of origin] 
    Intuitively, the client will only execute active content code (signified by the event
    $\CExecutes$ with $\ltextURL$ and \begin{nnew} manifest\end{nnew} 
    $\measurementInSignature$)
    if the code was uploaded by the honest developer $D$ (logged the event $\DUploads$),
    or the
    developer was corrupted.
    The $\KU$ event is emitted
    whenever the attacker (who is acting on behalf of the corrupted party $D$) constructs a message. 
    We simplify the formula as follows:%
    \begin{nnew}
    \begin{multline*}
        \CExecutes($\$D$, \, $\$\ltextURL$, \, \measurementInSignature)
        \implies
        \DUploads($\$D$, \, $\$\ltextURL$, \, \measurementInSignature)
        \; \lor \\
        (\Corrupted($\$D$) \land KU($\$\ltextURL$) \land KU(\measurementInSignature))
    \end{multline*}
    \end{nnew}
    Formally, the lemma is: for all $\CExecutes$ events there exists either an earlier $\DUploads$ event or 
    there exists a $\Corrupted($\$D$)$ event and $\KU$ events before $\CExecutes$ event.
\end{theorem}



\begin{theorem}[Transparency] 
    If the client
    executes \ac{JS} code $c$ for $\ltextURL$ with
    timestamp $\ts$ ($\CExecutes'$),
    then there is a corresponding log entry ($\Logged$) and
    it was deemed recent ($\CRecent$) by the client.
    The session identifier $\sid$ binds the moment when the client
    checks the timestamp is recent ($\CRecent$) to the moment it executes
    ($\CExecutes'$) the code.%
    \begin{nnew}
    \begin{multline*}
        \CExecutes'($\$\ltextURL$,\sid,c,\ts)
        \Rightarrow \\
            \Logged($\$\ltextURL$,c,\ts)
            \land
            \CRecent(\sid,\ts)
    \end{multline*}
    \end{nnew}
\end{theorem}
Authentication of origin and transparency describe the proactive
behaviour of the extension.
The following theorems cover the reactive behaviour. We first establish
that a claim that a client submits to the public is 
non-repudiable, i.e. that a corrupted client cannot forge false
evidence to implicate honest parties.

\begin{theorem}[Accountability]\label{theorem:claim-accept}
    When the public accepts a claim (identified with server id, $\ltextURL$, manifest, client nonce and log timestamp) 
    then, even if the client was corrupted, 
    the code must exist in the logs ($\Logged'$),
    and the server must have sent that data, either honestly, or
    dishonestly via the adversary.
    \begin{nnew}
    \begin{multline*}
        \PAccept($\$W$, \, $\$\ltextURL$, \, \measurementInSignature,n,ts)
        \implies
         \Logged'($\$\ltextURL$, \, \measurementInSignature,ts)
         \; \land \\
        (\LearnedFromW($\$W$, \, $\$\ltextURL$, \, \measurementInSignature,n) 
        \; \lor \\
         (\Corrupted($\$W$) \land \KU($\$W$, \, $\$\ltextURL$, \, \measurementInSignature,n))
    \end{multline*}
    \end{nnew}
    Here, the event $\LearnedFromW$ is emitted by $W$ (who is honest) right before it sends
    the signed tuple $\sigtuple$ to $C$ in 
    Fig.~\ref{fig:protocol-flow}.
\end{theorem}

\begin{theorem}[End to end guarantee]\label{theorem:e2e-guarantee}
When the client executes a malicious code, then a corrupted developer is necessary to distribute it.
    \begin{nnew}
    \begin{multline*}
            \CExecutes($\$D$, \, $\$\ltextURL$, \, `malicious`)
            \implies
             \Corrupted($\$D$)
    \end{multline*}
    \end{nnew}
\end{theorem}

\begin{theorem}[End to end non-guarantee]\label{theorem:e2e-nonguarantee}
When the client executes a malicious code, then a corrupted developer is sufficient to distribute it. 
    \begin{nnew}
    \begin{multline*}
            Ex.\; \CExecutes($\$D$, \, $\$\ltextURL$, \, `malicious`)
            \implies \\
             (All\; x.\; \Corrupted(x) \implies (x=$\$D$))
    \end{multline*}
    \end{nnew}
\end{theorem}

Tamarin reports these results within 3 hours 
on   
a 16-core computer with 2.6 GHz Intel Core i5 processors and 64 GB of RAM.
The proof is fully automatic, but relies on a so-called
`sources' lemma to specify were certain messages can originate from.
We specified this lemma manually, but it is verified automatically.
The full protocol can be found in the supplementary material~\cite{supplementary-files}.


%% file: claim-verification.tex
\section{Claim Verification}\label{sec:claimver}

The public runs a procedure to verify the claim 
generated by a client that was allegedly targeted by a website.
%
%
As shown in the Appendix~\ref{sec:protocol-verification} Theorem ~\ref{theorem:claim-accept},
a claim is identified with server name, URL, manifest, request nonce and the timestamp
that was set for the manifest by the ledger.
The signatures on the request and the response data are verified, and the request nonce is asserted with the server nonce for authenticity.
Next, the delivered content behaviours are checked against the manifest using the measurement procedure.
Then, the public evaluates if the manifest is the latest version on the ledger using the timestamp.
If the evaluation fails in any of these steps, then the claim is accepted.

%% file: eval-detail.tex
\section{Evaluation Details}\label{sec:evaldetail}
We provide details about evaluation of each case study as follows.

\subsection{`Hello World' Application Scenario}

    We evaluate the sample web application (\cref{html:deliver-manifest})
    using our browser extension.
    The manifest file in \cref{manifest:simple-example} is produced
    automatically.
    The extension fetches and verifies the signed 
    manifest linked in the response headers
    in parallel to collecting the metadata on the $\inline$ script.
    Signature validation, meta data collection and the final
    compliance check succeed with imperceptible impact on performance with Accountable JS.
    We evaluated the same scenario by enabling \ac{CSP} and disabling Accountable JS.
    We firstly deployed a detailed \ac{CSP} header that is close to Accountable JS manifests, 
    namely includes the valid sources for scripts and provides the hashes for the script and event handler resources. 

        \begin{lstlisting}[language=html,aboveskip=-1mm]

add_header content-security-policy "default-src 'self';script-src 'self' 'sha256-AfuyZ600rkX8AD+xANHUProHJm+22Tp0bMnvPFk/vas='; object-src 'none'"; 
\end{lstlisting} 
    
    We compared the \ac{CSP} results with the Accountable JS results and the difference is very small.

\subsection{Self-Contained Web Application Scenario}

    This case study shows that our browser extension is compatible with WhatsApp's web client.
    All active components are hosted on \url{web.whatsapp.com},
    it is thus easy to generate and maintain the manifest file.
    We observed that URLs for some external scripts include parts of their
    content's hash, likely related to caching optimisations.
    As any change to the content of an external script requires a new
    manifest file anyway, this is not a concern for our proposal.
    Moreover, $\generateManifest$ can automatically generate the
    manifest file.

    For measurement, we firstly mirrored the HTML page WhatsApp provides,
    locally to add integrity attributes for scripts. 
    However we observed that some dynamic behaviours 
    (e.g. script add and delete dynamically) did not exactly take place in local. 
    Hence, we evaluated the public website with
    integrity attributes for scripts stored hardcoded in the extension's content script
    (that collects metadata)
    and a hardcoded response header \texttt{x-acc-js-link} directed to a local URL that provides the signed manifest file.
    %
    %
    The WhatsApp Web application consists of
    nine \external and four \inline scripts present in the initial HTML.
    No more active content is appended after the window's load event,
    however, some external scripts are removed later.
    These removals occur during the measurement process,
    hence the automatically generated manifest successfully marks these
    external scripts with the $\persistent$ attribute and the others
    without it.
    The WhatsApp website is incompatible with Lighthouse tool,
    it shows a banner during the evaluation and the active contents are not delivered.
    Therefore, we use Puppeteer's Page metrics for evaluation.

    The traffic requirements are modest (about 1.17 bytes) and incur only modest blocking
    time. The performance overhead for combined duration of all tasks performed by the browser for baseline is 204ms, whereas
    it is 220ms for Code Verify and
    244ms for Accountable JS.
    The difference between Code Verify and Accountable JS is very small.
    This is remarkable, because
    Code Verify only applies \ac{SRI} checks on external scripts but not
    event handlers or iframes.
    In contrast to Accountable JS, the order of active elements is ignored, 
    attributes are not checked (e.g. \texttt{load='async'} for scripts) and
    a short hash value is downloaded from Cloudflare, rather than
    a signature.
%
    %

    %

\subsection{Trusted Third-party Code Scenario}
    
    This case study shows that our prototype is compatible with
    third party active contents and dynamic modification to the
    \ac{DOM}.
    Again, the extension captures the jQuery code and the
    inline \ac{JS}, and automatically generates a manifest.
    Note that the jQuery
    library removes the effectuating active content from the \ac{DOM},
    after interpreting the code and displaying the message in the
    window. It removes the script element in
    \cref{html:Trusted-third-party-code} and leaves only a
    non-executable message:

    \begin{lstlisting}[language=html]
<html><head>
  <script src="https://googleapis../jquery-3.6.1.min.js"></script>
</head>
<body>Hello World</body>
</html>
\end{lstlisting} 

    The measurement captures the inline script although it is
    removed and successfully compares the active content list against
    the manifest.
    The additional network traffic is almost identical to the simpler
    `Hello World' case. Likewise, the performance overhead is
    imperceptible.
    The difference between \ac{CSP} overheads and Accountable JS overheads is very small again.

\subsection{Delegate Trust to Third Parties Scenario}\label{sec:delegate-trust-eval-detail}

Nimiq's Wallet application behaves differently when the user does not have the Nimiq credentials stored in the browser and when the user has credentials.
It contains Hub in an iframe and Hub contains Keyguard in an iframe if the user does not have an account yet.
If the user has an account, the Hub does not embed the Keyguard.
We generated a manifest for Wallet that covers both cases by declaring the Keyguard with $\dynamic=true$.
In this way the Keyguard may or may not be delivered.
The browser extension will take the Keyguard into account if it is delivered; otherwise it will ignore it.
For this case study, we created a user account using Nimiq Hub prior the evaluation; hence the Keyguard is not embedded inside the Hub.

We created a simple shopping cart application
that uses Nimiq's Wallet in an iframe.
The first-party inline scripts communicate with the third party code
in the iframe via postMessages. As shown in
\cref{html:delegate-trusted-third-party-code}, the inline script
transmits a transaction record to the Nimiq Wallet.
In the main window's manifest (\cref{manifest:delegate-trust-third-party}),
the inline elements have $\mantrust=\assert$, while trust for the third party iframe (Wallet) is delegated to Nimiq.
Hence Nimiq's server is in charge of delivering a separate signed manifest for its content.
The main window's manifest was automatically generated, but we changed the
$\mantrust$ attribute for Wallet to $\delegate$.

The manifest for Nimiq's Wallet
(\cref{manifest:delegate-manifest-nimiq}) has seven \external scripts
and one iframe that is Hub.
We chose to assert trust for the external scripts via cryptographic
hashes.
For the Hub iframe, the manifest section has a nested manifest attribute that includes the manifest
sections for the active contents inside the Hub (six \external scripts, one \inline script and the iframe for the Keyguard).
%
For the KeyGuard iframe, being the most critical component, the manifest
section has also a nested manifest attribute that includes the manifest
sections for the active contents inside the KeyGuard
document.\footnote{We used both, for demonstrations in
    \cref{sec:use-cases}.
    $\generateManifest$ produces all variants, but the nested manifest
has precedence during the compliance checks.}
%
%
%
Our prototype successfully evaluates the active content list against
the manifests.
For efficiency, the prototype measures all elements 
simultaneously and combines their results once computed.
The browser extension panel displays results
instantaneously, while the evaluation is in progress, although the
evaluation is usually too quick for the user to notice.
The traffic requirements are about 9.9 kB for the additional
signature. In terms of performance, the impact of Accountable JS is again imperceptible.

In the \ac{CSP} evaluation for this case study, we defined \ac{CSP} headers for the main website, Wallet, Hub and Keyguard.
Accountable JS performance results are again close to \ac{CSP} except the total blocking time is slightly higher than the \ac{CSP}.
Besides, the reaction time unexpectedly decreases more with Accountable JS.
However, the difference in reaction time is minimal and could possibly be explained by
a) network latency,
(b) side effects of the browser's just-in-time compilation or
scheduling
or
(c) a side effect of the former two on how Lighthouse evaluates the
reactive metric.
Nimiq is a complex web application heavily dependent on external data,
in particular the remote blockchain it connects to.

\subsection{Untrusted Third-Party Code}

For Google AdSense, we tested with the active content of the framework, but no ads imported,
as we could not obtain an account with the provider.

As in the previous case,  we integrate the
high-security code (Nimiq's Wallet) as a sandbox and include
third party code AdSense with $\mantrust=\blindtrust$.
The web application, including AdSense, is functional and our
prototype shows compliance with the manifest. 
The network overhead is comparable to Nimiq A,
albeit slightly higher. This is due to the larger size of the
manifest, which now includes AdSense as well. 
Nevertheless, the performance
overhead is imperceptible with Accountable JS.

\subsection{Compartmentalisation of Code and Development Process}

As we stated before in Appendix~\ref{sec:delegate-trust-eval-detail}, 
Nimiq's Wallet behaves differently when the user does not have an account stored in the browser yet.
Furthermore, it redirects to Nimiq Hub from Wallet, when there is no account.
In this case study, we didn't create an account and 
since we have the local copy of Nimiq framework, we slightly updated the Nimiq's code to prevent the redirection,
to have the compartmentalisation case study (Wallet includes Hub in iframe and Hub further includes Keyguard in an iframe).

We evaluate the performance impact of compartmentalisation slightly
differently.
We consider Nimiq's Hub application, which includes the KeyGuard
application with $\mantrust=\delegate$ and thus requires a separate
manifest for the KeyGuard.
This time, a separate signing key is used for the KeyGuard manifest.
For the baseline performance, we inline the KeyGuard’s manifest as
an entry for its iframe in the Hub's manifest, thus having one
manifest and one signing key.
In contrast to the other case studies, the extension is activated in
the baseline measurement, too.

We observe that there are two more round trips and noticeably higher
traffic overhead (about the overhead of Accountable JS, not the
overall page traffic of 4,6 MB). 
This is due to downloading the extra \ac{SXG} certificate and manifest.
The effect on the rendering metrics is small, however there is a barely noticeable increase in reaction time.
Nevertheless, this can again be explained 
with network latency and side effects described above in Appendix~\ref{sec:delegate-trust-eval-detail}.

%% file: acronyms.tex
\section{Glossary}
\begin{acronym}[SXG] 
\acro{CA}{Certificate Authority}
\acro{CDN}{Content Delivery Network}
\acro{CSP}{Content Security Policy}
\acro{CT}{Certificate Transparency}
\acro{DOM}{Document Object Model}
\acro{JS}{JavaScript}
\acro{OCSP}{Online Certificate Status Protocol}
\acro{PKI}{Public Key Infrastucture}
\acro{SPA}{Single Page Applications}
\acro{SRI}{Subresource Integrity}
\acro{SXG}{Signed HTTP Exchanges}
\acro{TLS}{Transport Layer Security}
\acro{XSS}{Cross-Site Scripting}
\end{acronym}